\documentclass[fleqn,usenatbib]{mnras}

\usepackage{newtxtext,newtxmath,amsmath}

\usepackage[T1]{fontenc}

\DeclareRobustCommand{\VAN}[3]{#2}
\let\VANthebibliography\thebibliography
\def\thebibliography{\DeclareRobustCommand{\VAN}[3]{##3}\VANthebibliography}

\usepackage{graphicx,xcolor}	
\graphicspath{ {images/} }
\usepackage{amsmath}	
\usepackage{newtxtext,newtxmath}
\usepackage{cleveref}

\usepackage[mathscr]{euscript}
\DeclareSymbolFont{rsfs}{U}{rsfs}{m}{n}
\DeclareSymbolFontAlphabet{\mathscrsfs}{rsfs}

\usepackage{xcolor}

\title[The theory of kinks]{The theory of kinks --- I. A semi-analytic model of velocity perturbations due to planet-disc interaction}

\author[F. Bollati et al.]{
Francesco Bollati$^{1}$\thanks{E-mail: fbollati@uninsubria.it},
Giuseppe Lodato$^{1}$,
Daniel J. Price$^{2}$
and Christophe Pinte$^{2,3}$
\\
$^{1}$Dipartimento di Fisica, Universit\`a degli Studi di Milano, Via Celoria 16, Milano, Italy\\
$^{2}$School of Physics and Astronomy, Monash University, Clayton, Vic 3800, Australia\\
$^{3}$Universit\'e Grenoble Alpes, CNRS, IPAG, F-38000 Grenoble, France
}

\date{Accepted XXX. Received YYY; in original form ZZZ}

\pubyear{2021}
\defcitealias{Goodman01}{GR01}
\defcitealias{Rafikov02}{R02}

\begin{document}
\label{firstpage}
\pagerange{\pageref{firstpage}--\pageref{lastpage}}
\maketitle

\begin{abstract}
A new technique to detect protoplanets is by observing the kinematics of the surrounding gas. Gravitational perturbations from a planet produce peculiar `kinks' in channel maps of different gas species. In this paper, we show that
such kinks can be reproduced using semi-analytic models for the velocity perturbation induced by a planet. In doing so we i) confirm that the observed kinks are consistent with the planet-induced wake; ii) show how to quantify the planet mass from the kink amplitude; in particular, we show that the kink amplitude scales with the square root of the planet mass for channels far from the planet velocity, steepening to linear as the channels approach the planet; iii) show how to extend the theory to include the effect of damping, which may be needed in order to have localized kinks.



\end{abstract}




\section{Introduction}

Observations of planet-forming and planet-hosting discs have been revolutionized by the high spatial and spectral resolution of the Atacama Large Millimetre/submillimetre Array (ALMA) at mm-wavelengths. Initial observations in the dust continuum revealed significant substructures in the form of rings \citep{HLTau,Long18,Andrews18}, spirals \citep{Perez16,Huang18}, horseshoes \citep{Vandermarel13} and warps \citep[e.g.][]{Kraus20}. A common ---though not unique--- explanation for these structures is the dynamical interaction with either a stellar or sub-stellar companion, or a planet embedded in the disc. Rings and gaps in particular can be explained by the interaction of a Jupiter-mass planet with the dusty disc \citep{Dipierro15,Dipierro18,Veronesi20}. If planets are responsible for the creation of rings and gaps, the frequent occurrence rate of these substructures implies that massive planets may be already formed when the star is $\lesssim 1$ Myr old, meaning that planet formation occurs while star formation is ongoing. Explaining how a population of giant planets at tens of au separation can be already in place at such a young age challenges current models of planet formation \citep[e.g.][]{Zhang18,Lodato19}.

The above conclusion assumes that gaps and rings indicate planets --- an indirect inference. A direct connection between a gap and one (or actually two) planets came with the observation of the PDS70 system \citep{Keppler18,Keppler19}, where two directly imaged planets lie within a large dust gap.

More recently, line observations of gas species in the disc, such as the various CO isotopologues, have provided additional clues. Maps of the integrated line emission indicate that gas and dust distribution is not the same, with the dust disc smaller than the gas disc \citep{Ansdell18,Facchini19}, presumably related to radial drift of mm-sized grains (but see \citealt{Trapman19}). Line emission also allows one to probe the gas kinematics --- both small scale turbulence \citep{Flaherty15,Flaherty18,Flaherty20} and large scale ordered motion, e.g. Keplerian rotation of the disc.

Deviations from Keplerian motion offer a novel and effective way to find embedded planets, whose gravitational pull locally perturbs the gas kinematics. Such kinematical signatures of planets have been observed in four different ways. First, through the analysis of the rotation curve as derived from CO emission \citep{Teague18}, where a gap-opening planet reveals itself as a modification of the radial pressure profile, which in turn affect the rotation curve. Secondly, through the meridional circulation inside planetary gap, as laid out in \citet{Fung2016}, \citet{Dong2019} and \citet{Teague2019}. Thirdly, through the presence of a so-called `Doppler flip' in the moment 1 map of the gas \citep{Casassus19}.
Finally, one may search for characteristic ``kinks'' in individual channel maps, which appear in a limited number of channels and are spatially located close to the planet position \citep{Pinte18,Pinte19,Pinte20}. By comparing the ALMA data to hydrodynamical simulations, \citet{Pinte18,Pinte19,Pinte20} were able to constrain the mass of the planets responsible for the kinks. However, a detailed fitting procedure is made difficult by the need for 3D hydrodynamical simulations and the computational expense of fully exploring the parameter space.

\citet{GoldreichTremaine79,GoldreichTremaine80} and \citet{LinPap84} first developed the theory of the interaction between a gas disc and an embedded planet. The perturbing gravitational potential due to the planet excites density waves in the disc --- launched at resonant locations, the Lindblad resonances --- which propagate away from the planet, eventually becoming non-linear and steepening into shocks. \citet{Goodman01} and \citet{Rafikov02} (hereafter \citetalias{Goodman01,Rafikov02}) analysed the linear and non-linear density waves produced by a planet, considering mainly the density perturbations created by the wake. In principle, such a theory can be used to compute the velocity perturbations, too, and thus provide the theory needed to interpret line observations of planet-hosting discs. This is our aim in the present paper.

This paper is organized as follows: Section~\ref{theory} covers the basic theoretical framework of the model. In Section~\ref{sec:channelmaps} we derive the velocity perturbations due to planet-disc interaction and the corresponding channel maps. In Section~\ref{discussion} we discuss our results, with our conclusions presented in Section~\ref{conclusions}.

\section{Theoretical framework}
\label{theory}
\subsection{The shape of the wake}\label{wakeshape}
As pointed out by \cite{OgilvieLubow02}, the linear theory of planet disc interaction developed by \citet{GoldreichTremaine79, GoldreichTremaine80} does not provide information about the shape of the planet wake, since the disc response is determined separately for each Fourier mode of the planet potential. However, simulations for a terrestrial planet revealed the formation of a one-armed spiral wake \citep[e.g.][]{artymowicz2000}. \citet{OgilvieLubow02} explained this pattern as the result of constructive interference between density waves launched at different Lindblad resonances. They found that the planetary wake, in a frame corotating with the planet, lies on the curve
\begin{equation}
    \varphi _\textrm{wake}(r) =\varphi _\textrm{p} -\textrm{sgn}(r-r_\textrm{p})\frac{2}{3\varepsilon} \Biggl[\Bigl( \frac{r}{r_\textrm{p}} \Bigr)^{3/2}-\frac{3}{2}\ln \Bigl( \frac{r}{r_\textrm{p}} \Bigr) -1 \Biggr],
    \label{wakeeps}
\end{equation}
where $(r_\textrm{p},\varphi _\textrm{p})$ is the location of the planet in the disc in polar coordinates and $\varepsilon \equiv h(r)/r =$ const is the disc aspect ratio. By dropping the assumption of constant $\varepsilon$, the expression (\ref{wakeeps}) assumes the more general form
\begin{equation}
    \varphi _\textrm{wake}(r) = \varphi _\textrm{p} + \textrm{sgn}(r-r_\textrm{p})\int^r_{r_\textrm{p}}\frac{\Omega(r')-\Omega_\textrm{p}}{c_0(r')}\textrm{d}r',
    \label{wake}
\end{equation}
where $\Omega(r)$ and $\Omega _\textrm{p}$ are the disc and planet angular velocities and $c_0(r)$ is the unperturbed disc sound speed.
These quantities are related to the local disc thickness $h(r)$ by
\begin{equation}
    h(r) = \frac{c_0(r)}{\Omega(r)},
    \label{H}
\end{equation}
which holds for a thin disc in vertical hydrostatic equilibrium with a barotropic equation of state.
In the case of a Keplerian  power-law disc, where $\Omega (r) = (GM_\star/ r^{3})^{1/2}$ and $c(r) = c_\textrm{p}(r/r_\textrm{p})^{-q}$ with $q > 0$, Eq. (\ref{wake}) reads \citepalias{Rafikov02}
\begin{align}
    \varphi _\textrm{wake}(r) = \varphi _\textrm{p} + \textrm{sign}(r-r_\textrm{p}) &\biggl( \frac{h_\textrm{p}}{r_\textrm{p}}\biggr)^{-1}\biggl[ \frac{(r/r_\textrm{p})^{q - 1/2}}{q - 1/2} \nonumber \\
    &-\frac{(r/r_\textrm{p})^{q + 1}}{q + 1}-\frac{3}{(2q - 1)(q + 1)} \biggr].
    \label{wakepower}
\end{align}
The shape of the wake in the vicinity of the planet is found by
introducing local Cartesian coordinates $x=r-r_\textrm{p}$, $y = r_\textrm{p}(\varphi - \varphi _\textrm{p})$, expanding Eq. (\ref{wakeeps}) to the lowest non-vanishing order in $x/r_\textrm{p} \ll 1$ and using (\ref{H}).  The result, expressed in units of $(2/3)h_\textrm{p}$, reads
\begin{equation}
    y'_\textrm{wake} = -\textrm{sgn}(x')\frac{1}{2}x'^2,
    \label{parabola}
\end{equation}
where $x'=3x/2h_\textrm{p}$ and $y'=3y/2h_\textrm{p}$

Moreover, \cite{OgilvieLubow02} reported that the density perturbation of the wake has non-trivial internal structure, consisting of both a trough and a larger peak, as found from numerical simulations of planet-disc interaction.

\subsection{Linear perturbations in the shearing sheet approximation}\label{linear}
\citetalias{Goodman01} and \citetalias{Rafikov02} studied the internal structure of the density perturbation by solving the Euler and continuity equations of a disc with adiabatic equation of state semi-analytically. They split the computation of the density perturbation into two spatial regimes (where different approximations apply): a linear regime \citepalias{Goodman01} close to the planet and a nonlinear regime \citepalias{Rafikov02} further away from it.

In the linear regime the disc response is computed in the shearing sheet approximation, within a box of size $(8/3)h_\textrm{p}$ around the planet, where $h_\textrm{p}$ is disc thickness in correspondence of the planet.
The disc equations are written in local Cartesian coordinates $x$, $y$ in a corotating system centered on the planet. In this frame the planet wake is steady, so that the time derivatives of the perturbations vanish.
The fluid equations are i) linearized in the amplitude of small velocity and density perturbations and ii) the shearing sheet approximation is applied by assuming the homogeneity of the surface density $\Sigma _0$ and sound speed $c_0$ and by approximating the unperturbed disc motion to a linear shear flow.
The planet disturbance enters the equations as a point-like gravitational perturbation at the origin. No other effects related to its presence are considered.

By writing the resulting equations in Fourier space $(k_x, k_y)$, \citetalias{Goodman01} obtained an ordinary differential equation for the Fourier components of the azimuthal velocity perturbation $v$, while the Fourier components of the radial velocity perturbation $u$ and density perturbation $\sigma = \Sigma _1/\Sigma_0$ are given by algebraic relations (Equation~20 of \citetalias{Goodman01}).
For clarity, we report the basic equations here below:
\begin{equation}
    \begin{aligned}
    &\frac{\textrm{d}^2\hat{v}}{\textrm{d}\tau ^2}+\left[\kappa_\textrm{p} ^2 + c_\textrm{p}^2k^2(\tau)\right]\hat{v} = -ik_y\frac{\textrm{d}\hat{\Phi}_\textrm{p}}{\textrm{d}\tau} + 2ik_x(\tau)B_\textrm{p}\hat{\Phi}_\textrm{p}, \\
    &\hat{u} = - \frac{1}{c_\textrm{p}^2k_y^2+4B_\textrm{p}^2}\left[2iB_\textrm{p}k_y\hat{\Phi}_\textrm{p}+2B_\textrm{p}\frac{\textrm{d}\hat{v}}{\textrm{d}\tau}-c_\textrm{p}^2k_x(\tau)k_y\hat{v}\right], \\
    &\hat{\sigma} = \frac{i}{c_\textrm{p}^2k_y^2+4B_\textrm{p}^2}\left[ik_y^2\hat{\Phi}_\textrm{p}+k_y\frac{\textrm{d}\hat{v}}{\textrm{d}\tau} + \hat{v}k_x(\tau)2B_\textrm{p}\right]. \\
    \end{aligned}
    \label{linsystem}
\end{equation}
The Fourier transforms of the velocity and density perturbations $(\hat{v}, \hat{u}, \hat{\sigma})$ are functions of $\tau$ and $k_y$, where $\tau = -k_x/2A_\textrm{p}k_y$ and $A_\textrm{p} =(r\Omega '/2){|_{r_\textrm{p}}}$ is the first Oort constant at $r_\textrm{p}$. In Eqs. (\ref{linsystem}) we also see the second Oort constant $B_\textrm{p} = (A + \Omega){|_{r_\textrm{p}}}$, the epicyclic frequency $\kappa_\textrm{p} = (4\Omega B)^{1/2}{|_{r_\textrm{p}}}$ and the local sound speed $c_\textrm{p}\equiv c_0(r_\textrm{p})$ of the disc. The quantity $\hat{\Phi}_\textrm{p}$ in the r.h.s. of Eqs. (\ref{linsystem}) is the Fourier transform of the planet potential
\begin{equation}
    \hat{\Phi}_\textrm{p} = - \frac{2\pi G M_\textrm{p}}{k},
    \label{hatphi}
\end{equation}
where $M_\textrm{p}$ is the planet mass and $k = \sqrt{k_x ^2 + k_y ^2}$.
\citetalias{Goodman01} solved equations (\ref{linsystem}) numerically and computed $\sigma$ in real-space $(x,y)$ using a Fast-Fourier-Transform (FFT) algorithm. Following their numerical recipe, we solved the system (\ref{linsystem}) and transformed back to real-space to get also $u$ and $v$, as well as $\sigma$.

\begin{figure*}
    \centering
    \includegraphics[scale=0.3]{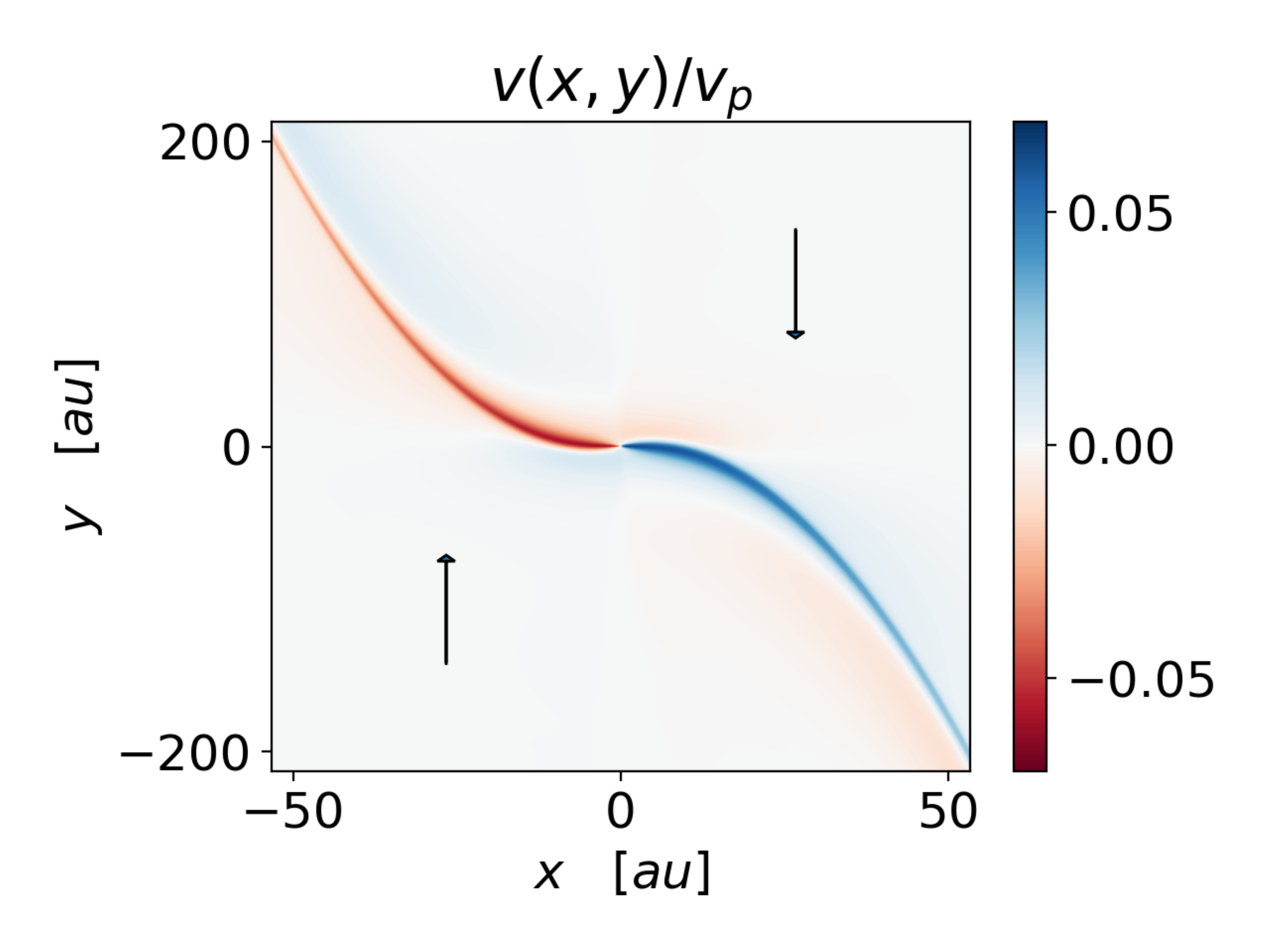}
    \includegraphics[scale=0.3]{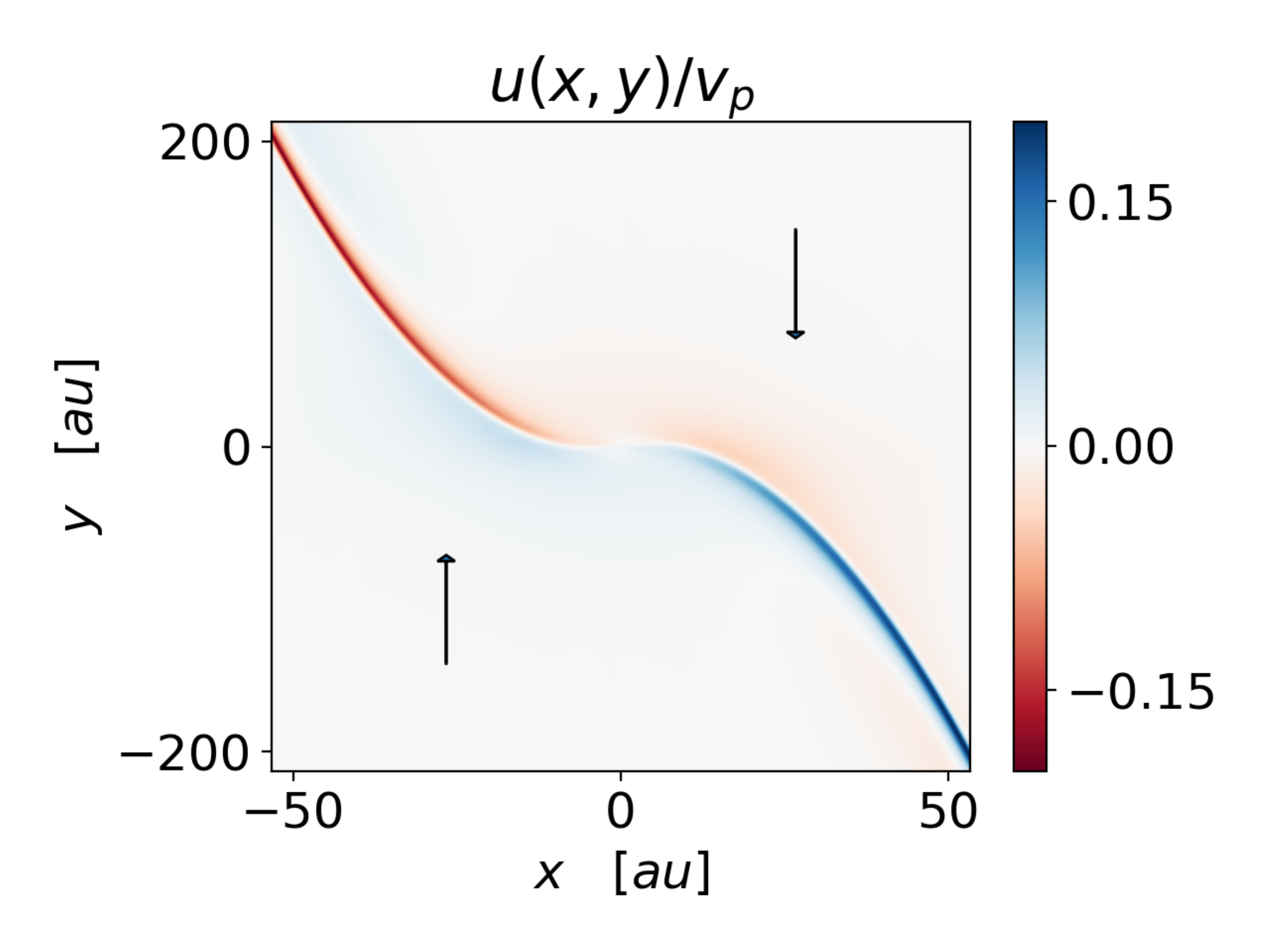}
    \caption{Linear azimuthal (left) and radial (right) velocity perturbations, in units of the planet Keplerian velocity $v_\textrm{p}$. Arrows show the direction of unperturbed disc flow in the local frame rotating with the planet.}
    \label{vlinear}
\end{figure*}

\begin{figure*}
    \centering
    \includegraphics[scale=0.3]{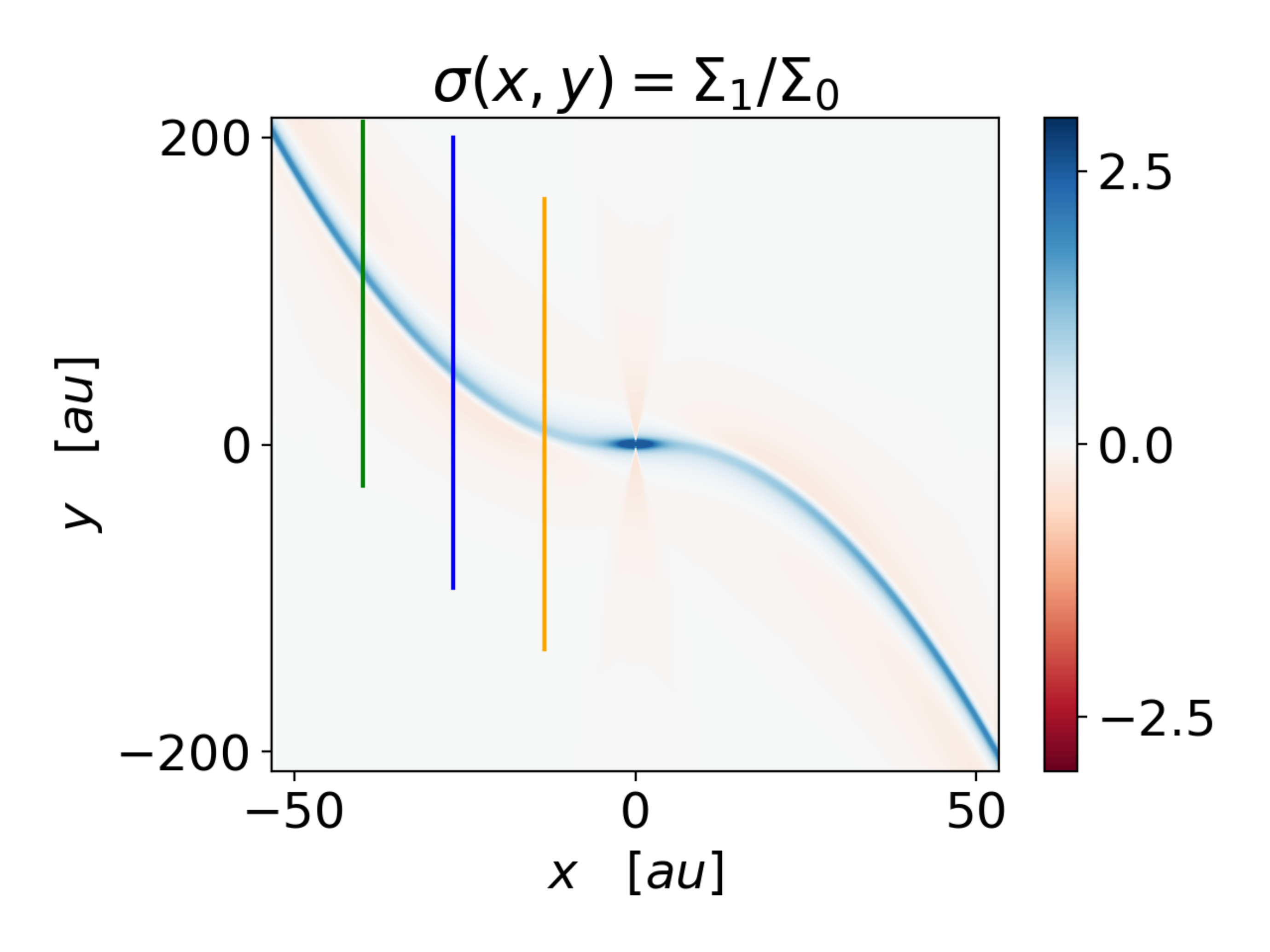}
    \includegraphics[scale=0.3]{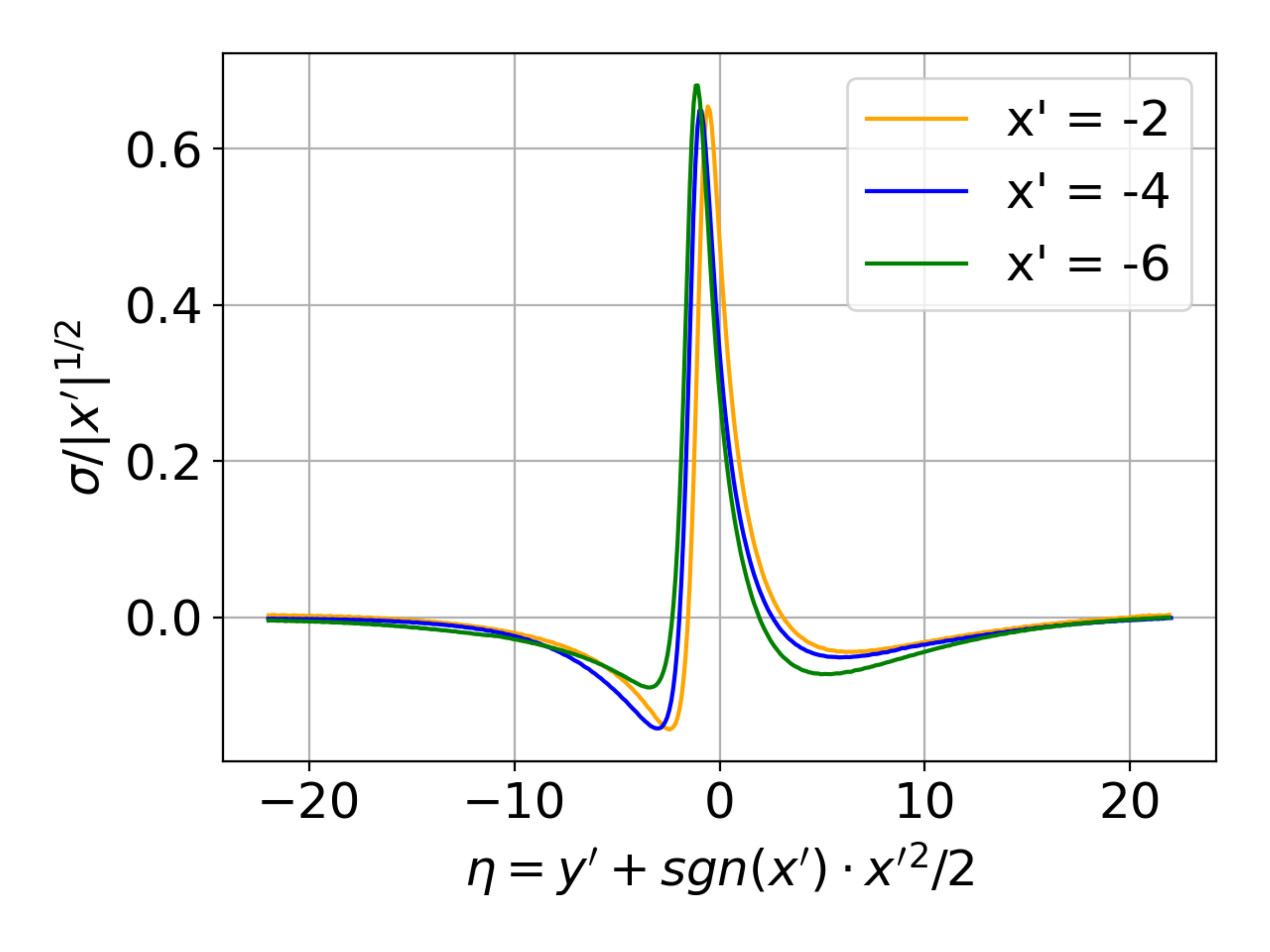}
    \caption{Left: Linear density perturbation in unit of local unperturbed disc surface density $\Sigma _0(r_\textrm{p})$. Right: Surface density profiles for different radii inside the planetary orbit corresponding to the three vertical lines shown in the left panel (Cf. figure 1 of \citetalias{Goodman01}). Primed coordinates are normalized with $(2/3)h_\textrm{p}$. The profile corresponding to $x'=-2$ is the initial condition of Eq. (\ref{burgers}) for $r<r_\textrm{p}$, modulo a scaling factor, as discussed in \cref{nonlinear} }
    \label{ss}
\end{figure*}

The computation is carried out with dimensionless quantities using $(2/3)h_\textrm{p}$ as length unit, $c_\textrm{p}$ as the unit of velocity and $m_\textrm{th}=(2/3)h_\textrm{p}c_\textrm{p}^2/G$ (Eq. 19 of \citetalias{Goodman01}) as the mass unit, which, thanks to (\ref{H}) and assuming Keplerian rotation, turns out to be the thermal mass
\begin{equation}
    m_\textrm{th} = \frac{2}{3}\left(\frac{h_\textrm{p}}{r_\textrm{p}}\right)^3 M_\star.
    \label{unitm}
\end{equation}
Definition (\ref{unitm}) implies that the scale of the linear domain is that of the Hill radius $r_\textrm{H}=r_\textrm{p}(m_\textrm{th}/3M_\star)^{1/3}$.
Using the units mentioned above and the fact that for Keplerian rotation $\Omega /2|A| = 2/3$ and $B/2|A| = 1/6$, the equation set (\ref{linsystem}) becomes
\begin{align}
    &\frac{\textrm{d}^2\hat{v}}{\textrm{d}\tau ^2}  +\biggl[k_y^2(\tau ^2+1) + \frac{4}{9}\biggr]\hat{v} =  - \textrm{sign}(k_y)\biggl(\frac{M_\textrm{p}}{m_\textrm{th}} \biggr)\frac{2\pi i}{3}\frac{\tau (\tau ^2+4)}{(\tau^2 + 1)^{3/2}}, \nonumber \\
    &\hat{u} = -\frac{1}{k_y^2+1/9}\biggl[ \frac{1}{3}\frac{\textrm{d}\hat{v}}{\textrm{d}\tau}-k_x(\tau)k_y\hat{v}-i\biggl(\frac{M_\textrm{p}}{m_\textrm{th}} \biggr)\frac{2\pi k_y}{3 k} \biggr], \nonumber \\
    & \hat{\sigma} = \frac{i}{k_y^2 + 1/9}\biggl[ k_y \frac{\textrm{d}\hat{v}}{\textrm{d}\tau} + \frac{1}{3}k_x(\tau)\hat{v} -i \biggl(\frac{M_\textrm{p}}{m_\textrm{th}} \biggr)\frac{2\pi k_y^2}{k}\biggr]. \label{dimensionless}
\end{align}
In order to solve these equations, we set the free parameter $M_\textrm{p}/m_\textrm{th}=1$ and
we considered a grid of $N_x \times N_y = 2^{12}\times 2^{13}$ points in $(k_x,k_y)$. Following \citetalias{Goodman01}, we fixed the maximum $k_y = 8$ and for each $k_y$ in the grid we solved the first equation of (\ref{dimensionless}) in the range $[-\tau_\textrm{max},\tau_\textrm{max}]$ using $\tau _\textrm{max} = (N_y \pi/8)^{1/2}$.
As the initial condition we used $\hat{v}=\textrm{d}\hat{v}/\textrm{d}\tau = 0$ at $\tau = - \tau _\textrm{max}$. Having obtained $(\hat{v}, \hat{u}, \hat{\sigma})$, we multiplied these quantities by a the pitch-angle filter, as described in \citetalias{Goodman01} \S 4, and finally transformed back to real-space via FFT algorithm.
This computation provided us the velocity perturbations, in unit of $c_\textrm{p}$, and the density perturbation, in unit of the local surface density $\Sigma _p \equiv \Sigma _0(r_\textrm{p})$, generated by a planet of mass $m_\textrm{th}$.

Figure~\ref{vlinear} shows the resulting velocity perturbations for a planet with mass equal to the thermal mass located at $r_\textrm{p}=100$ au embedded in a disc with
$h_\textrm{p}/r_\textrm{p}=0.1$, in orbit around a solar mass star ($M_\star = M_\odot$). With this choice of parameters, the
planet mass is $m_\textrm{th} \simeq 0.7 M_\textrm{Jupiter}$ and the linear box side measures $(8/3)h_\textrm{p} \simeq 26.67$ au.
The velocity perturbations are found to be anti-symmetric w.r.t the origin, as expected from the symmetries of
Eqs. (\ref{dimensionless}) and the properties of the Fourier transform, and lie on the curve (\ref{parabola}), in agreement with \cite{OgilvieLubow02}.

Figure~\ref{ss} illustrates the corresponding density perturbation, reproducing the result of \citetalias{Goodman01}.

In this regime, all perturbations are linear in the planet mass, as it emerges from equations (\ref{dimensionless}), therefore the perturbations for a generic planet mass $M_\textrm{p}$ can be found with respect to a planet of mass $m_\textrm{th}$ by rescaling the perturbations by a factor $(M_\textrm{p}/m_\textrm{th})$. Note that the linear approximation holds only for planet masses $\ll m_\textrm{th}$ that do not open a gap in the disc.

\subsection{Global, non-linear evolution}\label{nonlinear}

We now turn to the structure of the density perturbation outside the linear domain, as first analysed by \citetalias{Rafikov02}. We summarize here some relevant aspects.
In \citetalias{Rafikov02}, the density perturbation at the edges of the linear domain was used as the initial condition for the computation of the density perturbation away from the planet. In this regime, i) the shearing sheet approximation is abandoned, recovering the radial dependence of the disc structure, and ii) the disc equations are simplified via the tight-winding approximation. In addition, iii) the planet source term is dropped and iv) second order nonlinear terms are considered.
Under these assumptions, \citetalias{Rafikov02} showed that the structure of the density perturbation, in a polar coordinate frame corotating with the planet\footnote{In this frame the perturbation is stationary, as in the linear regime.}, obeys the inviscid Burgers' equation\footnote{In \citetalias{Rafikov02}'s equation for $\chi$, the sgn($r-r_\textrm{p}$) is replaced by a minus sign since only the case $r<r_\textrm{p}$ is considered. The symmetry of $\sigma$ suggests that in the general case a sgn($r-r_\textrm{p}$) term must be present, as confirmed by following \citetalias{Rafikov02}'s derivation of the equation for $\chi$ (Appendix A of \citetalias{Rafikov02}) in the case $r>r_\textrm{p}$.}
\begin{equation}
    \partial _t \chi +\textrm{sgn}(r-r_\textrm{p}) \chi \partial _{\eta}\chi = 0,
    \label{burgers}
\end{equation}
where $\chi$ is related to the density perturbation $\Sigma - \Sigma_0$ via
\begin{align}
    &\chi \equiv \frac{\gamma + 1}{2}\frac{\Sigma -\Sigma _0}{\Sigma _0}g(r), \label{chiii} \\
    &g(r) \equiv \frac{2^{1/4}}{r_\textrm{p} c_\textrm{p} \Sigma _\textrm{p} ^{1/2}}\biggl(\frac{r\Sigma _0 c_0^3}{|\Omega - \Omega _\textrm{p}|} \biggr)^{1/2}, \label{ggg}
\end{align}
$\gamma$ is the disc adiabatic index and subscript `p' denotes unperturbed quantities evaluated at $r_\textrm{p}$.
The coordinates $t$ and $\eta$ in Equation~(\ref{burgers}) are spatial coordinates defined by
\begin{align}
    &t(r) \equiv -\frac{r_\textrm{p}}{2h_\textrm{p}/3}\int _{r_\textrm{p}}^r\frac{\Omega (r')-\Omega _\textrm{p}}{c_0(r')g(r')}\textrm{d}r', \label{t} \\
    &\eta(r,\varphi) \equiv \frac{r_\textrm{p}}{2h_\textrm{p}/3}[\varphi - \varphi _\textrm{wake}(r)], \label{eta1}
\end{align}
where $\varphi _\textrm{wake}$ is given by (\ref{wake}).
Eqs. (\ref{t}) and (\ref{eta1}) can be thought of, respectively, as a coordinate along the spiral wake and as an azimuthal coordinate centered in the wake at the same radius. Hence, Eq. (\ref{burgers}) describes the ``evolution'' of the azimuthal profile of the density perturbation along the spiral wake. For the case of a Keplerian power-law disc ($\Sigma_0(r) = \Sigma _\textrm{p}(r/r_\textrm{p})^{-\delta}$ , Eq.~\ref{wakepower} and preceding text), the expression for $g(r)$ (Eq.~\ref{ggg}) becomes
\begin{equation}
    g(r) = 2^{1/4} \left(\frac{h_\textrm{p}}{r_\textrm{p}}\right)^{1/2} \frac{(r/r_\textrm{p})^{5/4 - (\delta + 3q)/2}}{ \vert 1  - (r/r_\textrm{p})^{3/2} \vert^{1/2}},
\end{equation}
giving the integral for $t(r)$ (Eq.~\ref{t}) in the form
\begin{equation}
    t(r) = \frac{-3}{2^{5/4}(h_\textrm{p}/r_\textrm{p})^{5/2}} \int_1^{r/r_\textrm{p}} (1 - x^{3/2}) \vert 1 - x^{3/2} \vert^{1/2} x^w dx,
\end{equation}
where $w \equiv -11/4 + (\delta + 5q)/2$ and we substituted $x \equiv r/r_\textrm{p}$.  Since $t$ is always positive one may also simplify further to
\begin{equation}
    t(r) = \left\vert \frac{-3}{2^{5/4}(h_\textrm{p}/r_\textrm{p})^{5/2}} \int_1^{r/r_\textrm{p}} \vert 1 - x^{3/2} \vert^{3/2} x^w dx \right\vert ,
\end{equation}
which is the integral we solve in the code.

In order to solve Eq. (\ref{burgers}), we need to specify the initial condition. We observe that Eq. (\ref{burgers})
consists of two equations, one for $r>r_\textrm{p}$ and the other for $r<r_\textrm{p}$, hence two initial profiles, $\chi_+(\eta)$ and $\chi_-(\eta)$, are required. These functions are found by taking the limit of $\chi$ in the linear regime and by evaluating the result at the edges
$r_\pm=r_\textrm{p} \pm (4/3)h_\textrm{p}$ of the linear box domain.
More precisely, in linear and shearing sheet approximation, $\chi$ reduces to (cf. Eq. 35 of \citetalias{Rafikov02})
\begin{equation}
    \chi(t,\eta) \approx \frac{\gamma + 1}{2^{3/4}}\frac{M_\textrm{p}}{m_\textrm{th}}\frac{\sigma\bigl(x',\eta-x'^2\textrm{sgn}(x')/2\bigr)}{\sqrt{x'}}
    \label{chinear}
\end{equation}
where $\sigma = (\Sigma-\Sigma_0)/\Sigma _0$ is the density perturbation\footnote{We use the same symbol $\sigma$ for both the density perturbation as a function of $x$, $y$ and as a function of dimensionless $x'$, $y'$, where it is implicit that $\sigma(x,y) = \sigma (x',y')$.
} computed in the linear regime for the thermal mass $m_\textrm{th}$, the factor $M_\textrm{p}/m_\textrm{th}$ is due to the linear scaling of $\sigma$ with $M_\textrm{p}$ and the second argument of $\sigma$, $y'$, has been written using $\eta \simeq y' - y'_\textrm{wake}$, which is the form that Equation~(\ref{eta1}) assumes near the planet.

 Figure~\ref{ss} (right panel) shows the corresponding profiles of (\ref{chinear}) for some example values of $x'$, modulo the prefactor $(M_\textrm{p}/m_\textrm{th})(\gamma +1)2^{-3/4}$, for the case $r<r_\textrm{p}$.
Since $\sigma$ is symmetric with respect to the origin, the profiles of (\ref{chinear}) for $r>r_\textrm{p}$ are simply flipped with respect to the $y$ axis compared to those shown in Figure~\ref{ss} (right panel).
Then, the initial conditions $\chi _\pm$ of the problem (\ref{burgers}) are given by (\ref{chinear}) evaluated at $x' = (r_\pm -r_\textrm{p})/(2h_\textrm{p}/3) =\pm 2$, the outer and inner boundaries of the linear domain. This corresponds to set the initial ``time'' of nonlinear ``evolution'' at  $t_0 \simeq 1.89$.


Having computed the initial conditions for $\chi$, we proceed to discuss its evolution under Equation~(\ref{burgers}).

\begin{figure}
    \centering
    \includegraphics[scale=0.35]{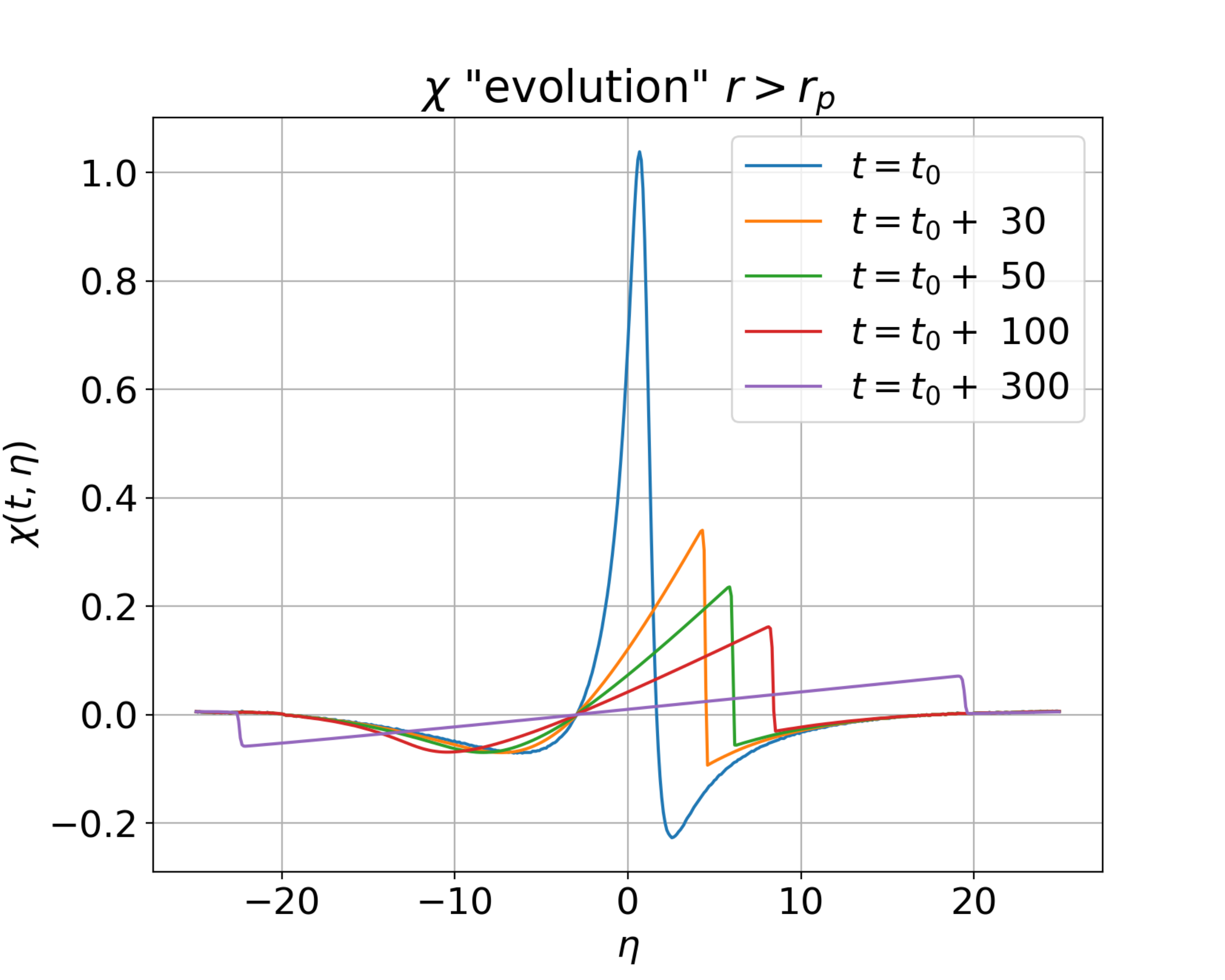}
    \caption{Time evolution of the numerical solution $\chi$ of Eq. (\ref{burgers}) for $r>r_\textrm{p}$ in case of $M_\textrm{p} = m_\textrm{th}$.
    The wave elements move to the right at speed
    $(\partial \eta / \partial t)_\chi = \chi$.
    In this way the initial negative hump at $\eta<0$ moves to the left and shocks, preserving the initial area of the hump. On the other hand, the negative hump at $\eta>0$, is absorbed into the shock moving to the right formed by the positive peak, whose area is reduced by that of the incorporated negative hump.
    Asymptotically $\chi$ develops an N-wave profile.}
    \label{ichievolution}
\end{figure}

From the theory of Burgers' equation (\citealt{whitham74}) we know that each wave element of an initial profile travels with a characteristic speed that depends on the value of $\chi$ carried by that the wave element.
More formally, a certain value $\chi$ is kept constant along the characteristic curve $(t,\eta(t))$ of speed $(\partial \eta / \partial t)_\chi = \textrm{sgn}(r-r_\textrm{p})\chi$,
since on such curve $\textrm{d}\chi/\textrm{d}t = 0$, as follows from Equation~(\ref{burgers}). In particular, high-$\chi$ elements move faster than low-$\chi$ ones and, as a consequence, the wave profile is distorted and eventually shocks (see Figure~\ref{ichievolution}).
As shown in \cite{whitham74} Ch. 2.8, if the initial waveform changes sign and has a finite range (as the profiles in Figure~\ref{ss} do), an ``N-wave'' shape eventually develops, with two shocks of opposite signs moving in opposite directions. The areas of the two lobes of the N-wave are constant in time,
since the flux across the point $\chi=0$ that separates the lobes vanishes and the ``mass''
$\int_{\textrm{annulus}} \chi\textrm{d}\eta$ is conserved. In our case, the lobe areas are equal to each other because the conserved ``mass'' is zero\footnote{
This is true because for every $t\geq t_0$, that is for every radius both $r\geq r_+$ and $r\leq r_-$, the integral of $\chi$ on the whole range of $\eta$, a disc annulus, must vanish $\int \chi\textrm{d}\eta \propto \int_0^{2\pi}(\Sigma - \Sigma _0)\textrm{d}\varphi = 0$, since the wake density perturbation alters the azimuthal density profile of the annulus but not its mass.}. Due to their time independence, the areas of the lobes can be found from the profile at $t_0$. In our problem for $r<r_\textrm{p}$, such areas are given by
\begin{equation}
 A\equiv \Bigl|\int_{\widetilde{\eta}}^{\infty}\chi_-(\eta)\textrm{d}\eta\Bigr|,
 \label{A}
\end{equation}
where $\widetilde{\eta} \simeq 2.96$ is the point $\chi_-(\widetilde{\eta})=0$ separating the lobes (see Fig. \ref{ss}). It turns out that the lobes areas for $r>r_\textrm{p}$ are given by (\ref{A}) as well because of the symmetry of $\sigma$.

Concerning the dependence on the planet mass, we know that the height and width of the lobes scale as $A^{1/2}(t-t_0)^{-1/2}$ and $A^{1/2}(t-t_0)^{1/2}$ (\citealt{whitham74}), hence these quantities are both proportional to $M_\textrm{p}^{1/2}$, as it follows from (\ref{A}) and  $\chi _-\propto \sigma \propto M_\textrm{p}$.

The asymptotic N-wave solutions of (\ref{burgers}) can be written in a form which comprises the two cases $r\lessgtr r_\textrm{p}$. This formula reads:
\begin{equation}
    \chi (t,\eta) =
    \begin{cases}
       \left[\textrm{sgn}(r-r_\textrm{p})\eta + \widetilde{\eta}\right]/(t-t_0) &\eta \in [\eta _-,\eta_+] \\
       0 & \textrm{elsewhere}
    \end{cases}
    \label{Nwave}
\end{equation}
where
\begin{equation}
    \eta _\pm = -\textrm{sgn}(r-r_\textrm{p})\widetilde{\eta}\pm\sqrt{2A(t-t_0)}.
    \label{etapm}
\end{equation}
This solution of Equation~(\ref{burgers}) applies asymptotically, i.e. far away from the planet. In order to compute the density perturbation near the planet we need to find a numerical solution of Equation~(\ref{burgers}). We solved Equation~(\ref{burgers}) using a Godunov scheme on a grid of $N_\eta \times N_t = 509 \times 14966$ points.

 Figure~\ref{ichievolution} shows the time evolution of the solution (recalling that `time' here is a coordinate related to distance along the wake) for the case $r>r_\textrm{p}$. The solution for $r<r_\textrm{p}$ can be found noting that if $\chi_\textrm{out}$ is the solution of (\ref{burgers}) for $r>r_\textrm{p}$, then $\chi_\textrm{in}(t,\eta) \equiv \chi _\textrm{out}(t,-\eta)$ satisfies (\ref{burgers}) for $r<r_\textrm{p}$.
From Figure~\ref{ichievolution}, in which $M_\textrm{p} = m_\textrm{th}$, we see that after a time $\simeq t_\infty \equiv t_0 + 300$, the profile displays a N-wave shape and therefore it can be described with the asymptotic N-wave solution (\ref{Nwave}).
For a generic planet mass, the time $t_\infty$ after which the profile is well approximated by the N-wave is $t_\infty = t_0 +300 (m_\textrm{th}/M_\textrm{p})$, according to Eq. (39) in \citetalias{Rafikov02}.

\section{Velocity perturbations and channel maps}
\label{sec:channelmaps}
While the spiral density perturbation resulting from planet-disc interaction is well understood, the associated velocity perturbations have not received the same attention. But this is needed in order to interpret the kinematical signatures of embedded planets!

\subsection{Velocity perturbations in nonlinear regime}\label{nlvelocity}
To produce the kinks seen in the channel maps,  we first need to compute the disc velocity field in the presence of a planet. We already achieved this in the vicinity of the planet in \cref{linear}. Here we complete the calculation by computing the nonlinear velocity perturbations, providing their relation to the density perturbation $\chi$. In doing so, we rely on the work of \citetalias{Rafikov02}, keeping the same notation.

As shown in section 2.2 of \citetalias{Rafikov02}, the nonlinear radial velocity perturbation is\footnote{Actually \citetalias{Rafikov02} presents $u$ without the coefficient $-\textrm{sgn}(r-r_\textrm{p})$,
since he consideres only the case $r<r_\textrm{p}$. By following \citetalias{Rafikov02}'s derivation of $u$ (section 2.2 of \citetalias{Rafikov02}) for $r>r_\textrm{p}$ we easily obtain Equation~ (\ref{urafikov}).}
\begin{equation}
    u = -\textrm{sgn}(r-r_\textrm{p})\frac{2(c_0-c)}{\gamma - 1}.
    \label{urafikov}
\end{equation}
This equation can be rewritten in terms of the quantity $\psi = (c-c_0)(\gamma + 1)/[c_0(\gamma - 1)] \ll 1$ (Equation~A2 of \citetalias{Rafikov02}) to yield
\begin{equation}
    u = \textrm{sgn}(r-r_\textrm{p})\frac{2c_0}{\gamma + 1}\psi.
    \label{upsi}
\end{equation}
On the other hand, the azimuthal velocity perturbation, expressed as a function of $\psi$, reads (see Equation~A8 of \citetalias{Rafikov02})
\begin{equation}
    v \simeq - 2\frac{c_0^2}{\Delta \Omega r}\frac{1}{\gamma + 1}\psi.
    \label{vpsi}
\end{equation}
where $\Delta \Omega \equiv \Omega - \Omega _\textrm{p}$.
Now we write $u$ and $v$ in terms of $\chi$ using the relation
\begin{equation}
    \chi = g\psi + \mathcal{O}(\psi^2),
    \label{chipsi}
\end{equation}
which follows from Equation~(A3) of \citetalias{Rafikov02} and definition (\ref{chiii}).
By replacing (\ref{chipsi}) in (\ref{upsi}) and (\ref{vpsi}), discarding $\mathcal{O}(\psi ^2)$ terms\footnote{As in \citetalias{Rafikov02}'s derivation of Eq. (\ref{burgers}), where nonlinear
terms proportional to $\psi \partial _r \psi$ are retained while $\psi ^2$ terms are discarded, since the tidal perturbation is assumed to be weak and tightly wound.} and using Equation~(\ref{ggg}), we obtain the nonlinear velocity perturbations
\begin{align}
    u(r,\varphi) &= \textrm{sgn}(r-r_\textrm{p})\Lambda _\textrm{p} f_u(r) \chi (t,\eta) \label{unl}, \\
    v(r,\varphi) &= \textrm{sgn}(r-r_\textrm{p})\Lambda _\textrm{p} f_v(r) \chi (t,\eta) \label{vnl},
\end{align}
where
\begin{align}
    &\Lambda _\textrm{p} = \frac{2^{3/4}}{\gamma + 1}r_\textrm{p} c_\textrm{p}\Sigma _\textrm{p}^{1/2}, \label{Lambda} \\
    &f_u(r) = \left[ \frac{|\Delta \Omega (r)|}{r\Sigma _0(r)c_0(r)} \right]^{1/2},  \label{fu}\\
    &f_v(r) = \left[ \frac{c_0(r)}{|\Delta \Omega (r)|\Sigma _0(r)r^3} \right]^{1/2}  \label{fv},
\end{align}
and $\chi = \chi\left(t(r),\eta (r,\varphi); M_p\right)$ is the solution to Equation~(\ref{burgers}). The term (\ref{fv}) formally diverges at $r=r_\textrm{p}$ but we need not worry since equations (\ref{unl})-(\ref{fv}) hold in the nonlinear regime, outside the linear box centered in $r_\textrm{p}$ where the singularity occurs.
If the disc is a power-law Keplerian disc, with surface density and sound speed profiles $\Sigma_0(r)=\Sigma_\textrm{p}(r/r_\textrm{p})^{-\delta}$ and $c_0(r)=c_\textrm{p}(r/r_\textrm{p})^{-q}$, the terms $\Lambda _\textrm{p} f_u(r)$ and $\Lambda _\textrm{p} f_v(r)$ read
\begin{align}
    \Lambda _\textrm{p} f_u(r) &= c_\textrm{p} \left(\frac{h_\textrm{p}}{r_\textrm{p}}\right)^{-1/2}\frac{2^{3/4}}{\gamma + 1}\left( \frac{r}{r_\textrm{p}}\right)^{(\delta + q - 1)/2}\left| \left(\frac{r}{r_\textrm{p}}\right)^{-3/2}-1\right|^{1/2}, \label{pwlu} \\
    \Lambda _\textrm{p} f_v(r) &= c_\textrm{p} \left(\frac{h_\textrm{p}}{r_\textrm{p}}\right)^{1/2}\frac{2^{3/4}}{\gamma + 1}\left( \frac{r}{r_\textrm{p}}\right)^{(\delta - q - 3)/2}\left| \left(\frac{r}{r_\textrm{p}}\right)^{-3/2}-1\right|^{-1/2}, \label{pwlv}
\end{align}
where we also used Equation~(\ref{H}).

From Equations~(\ref{unl})--(\ref{fv}) it follows that the azimuthal profiles of $u$ and $v$ have the same shape of $\chi$ --- rescaled by the factors $\Lambda _\textrm{p}f_u(r)$ and $\Lambda _\textrm{p}f_v(r)$. Therefore, asymptotically, they display a N-wave shape
with the same scaling properties $\propto M_\textrm{p}^{1/2}$ of $\chi$, as discussed in \cref{nonlinear}.

The velocity perturbations (\ref{unl})--(\ref{fv}) in addition to those computed in the linear regime (\cref{linear}) represent the velocity disturbance caused by the planet wake. We assume that this is the planet-induced velocity substructure responsible for the observed kinks, although other origins have been suggested \citep[e.g.][]{Perez15}. Under this assumption, in the following sections we attempt to produce analytical kinks from the velocity perturbations of the wake and compare them to the observations.

\subsection{Channel maps in a Keplerian disc} \label{channels}

Before delving into the kinks, let us consider the line emission expected in a channel map in the absence of a planet, i.e. when the disc velocity field is unperturbed. We assume Cartesian coordinates such that the origin coincides with the star, the disc lies in the $xy$-plane and the $z$ axis forms an angle $\theta$ with the line-of-sight $\hat{n}$ (pointing towards the observer). In addition, the $x,y$ axes are oriented such that  $\hat{n}=(-\sin\theta,0,\cos\theta)$,
which corresponds to a disc with PA = 90$^{\circ}$ and inclination $i = \theta$.

\begin{figure}
    \centering
    \includegraphics[scale=0.3]{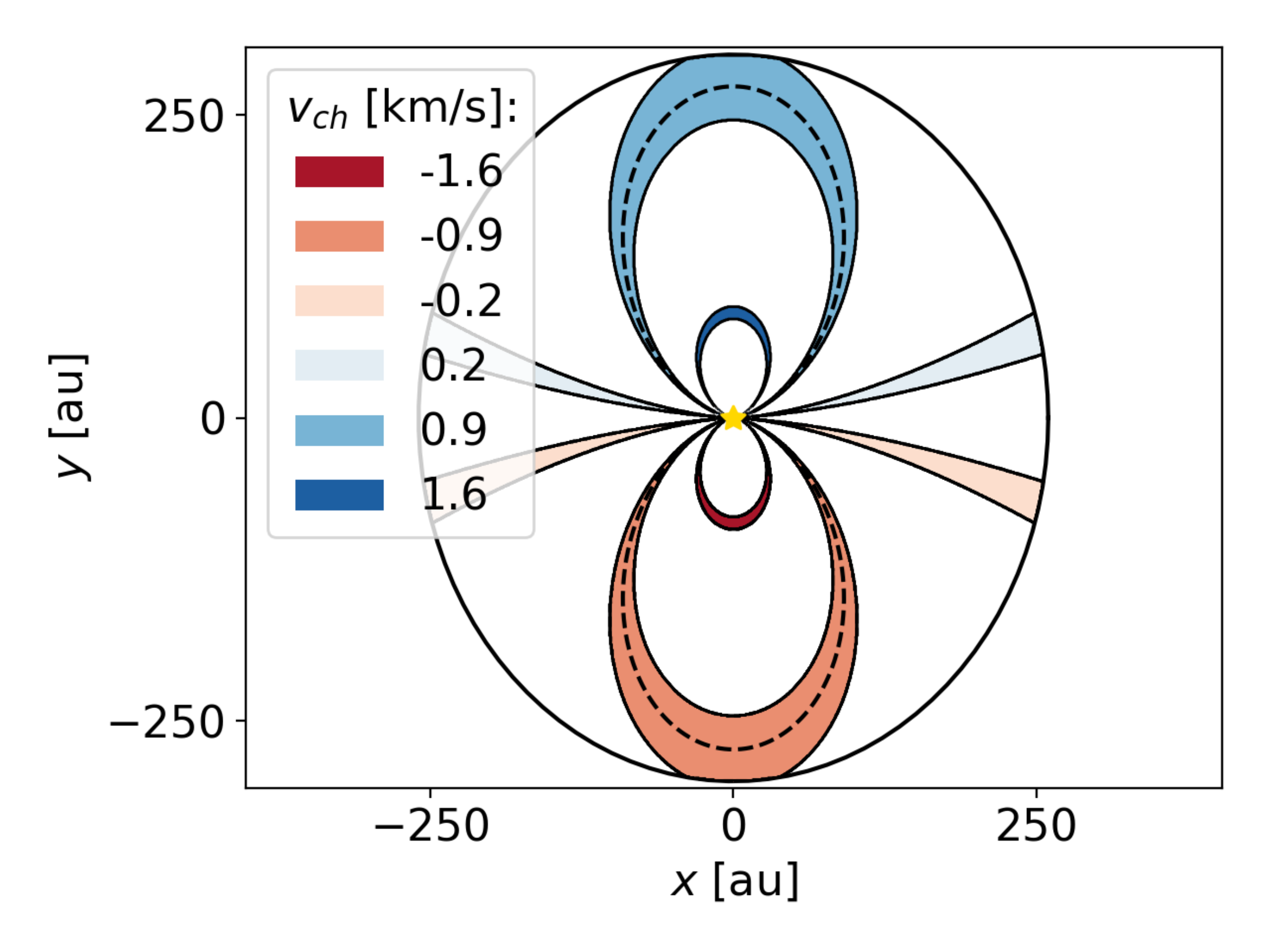}
    \caption{Channel maps for a Keplerian disc with inclination angle $\theta = \pi/6$ and velocity resolution $\Delta v = 0.05$ km/s. The dashed lines are examples of channel maps plotted from Eq. (\ref{anaiso}).}
    \label{iCM1}
\end{figure}

If the disc motion is undisturbed and inviscid, the only non-vanishing component of the disc velocity field is the azimuthal component. Moreover, if we neglect terms of order $\mathcal{O}((h/r)^2)$, this component can be approximated with the Keplerian velocity $v_\textrm{K} = (GM_\star/r)^{1/2}$. In this case we write the unperturbed disc velocity field as\footnote{In this notation the disc flows in counterclockwise direction as seen from $z$. A disc in clockwise rotation w.r.t. the observer corresponds to an inclination angle $\theta \in [\pi/2,3\pi/2]$.}
\begin{equation}
    \mathbf{v}_0 = v_\textrm{K}(r)\mathbf{e}_{\varphi}
    = v_\textrm{K}(r)(-\sin \varphi,\cos \varphi,0)^{\textrm{T}}.
    \label{velocityd}
\end{equation}
Molecular-line observations probe the line-of-sight component of the gas which, for a Keplerian disc, assumes the form\footnote{Here we assume that the disc is 2D, hence a given line of sight only penetrates the disc at a given $(r,\varphi)$. In reality if we consider a 3D structure, a given line of sight goes through multiple regions in the disc with different $(r,\varphi)$, implying a velocity projection different from Eq. (\ref{vnk}).}:
\begin{equation}
    v_{n,0} = \mathbf{v}_0\cdot \hat{n} =v_\textrm{K}(r)\sin \theta \sin \varphi.
    \label{vnk}
\end{equation}
For a generic velocity field $\mathbf{v}$, the  function $v_{n}(r,\varphi) = \mathbf{v}(r,\varphi)\cdot \hat{n}$ allow us to produce the corresponding channel maps. If the disc elements have a certain velocity field with non-zero component towards the observer ($v_n \neq 0$), then the emission will be Doppler-shifted. In a given channel map the emission is distributed along the regions of the source
that undergo the same Doppler-shift, i.e. the regions where
$v_n$ is equal to the channel velocity $v_\textrm{ch}$. Then, the disc elements $(r,\varphi)$ of a channel map, or isovelocity curve, with channel velocity $v_\textrm{ch}$ must satisfy the condition
\begin{equation}
 v_n(r,\varphi) = v_\textrm{ch}.
 \label{eqchannel}
\end{equation}
If the disc is Keplerian, $v_n$ is given by (\ref{vnk}) and Equation~(\ref{eqchannel}) admits the solution
\begin{equation}
    \widetilde{r}(\widetilde{\varphi}) = \frac{GM_\star}{v_\textrm{ch}^2}\Biggl| \frac{\cos \theta}{\cos \widetilde{\varphi}}\Biggr|\frac{(\sin \theta \cos \theta \tan \widetilde{\varphi})^2}{\bigl[ 1 + (\tan \widetilde{\varphi} \cos \theta)^2 \bigr]^{3/2}}\mathscrsfs{H}(\sin \widetilde{\varphi} \sin\theta v_\textrm{ch}),
    \label{anaiso}
\end{equation}
where $\mathscrsfs{H}$ is the Heaviside step function and $\widetilde{r}$, $\widetilde{\varphi}$ are the polar coordinates in the plane perpendicular to the line-of-sight (the sky plane), with $y$-axis coinciding with the disc inclination axis (see Figure~\ref{iCM1}).
Actually, in real measurements we always have a finite channel resolution $\Delta v$ (which is typically of the order of 0.05 km/s), therefore the channel maps will be given by the set of disc elements such that
\begin{equation}
   v_n(r,\varphi) \in [v_\textrm{ch}-\Delta v,v_\textrm{ch}+\Delta v].
   \label{channn}
\end{equation}
Figure \ref{iCM1} illustrates some examples of channel maps for a disc with a radius of 300 au in Keplerian rotation around a solar mass star (similar to \citealt{Pinte19}), considering different channel velocities and an inclination angle $\theta = \pi/6$. In this Figure the channel maps have been created by ``painting'' the grid points of the disc\footnote{In this and other Figures, we discretised the disc using a Cartesian grid of $10^3 \times 10^3$ points.} that satisfy the condition (\ref{channn}). The patterns shown in Fig. \ref{iCM1} are known as \emph{butterfly patterns} and are characteristic of discs in Keplerian rotation \citep[e.g.][]{Horne86}. In principle, a disturbance in the disc velocity field could be detected as a deviation from the butterfly pattern in the channel maps.
The channel maps which are coloured in tones of red represent red-shifted molecular lines, while the channel maps in tones of blue are associated with blue-shifted emissions. The red-blue shift patterns are reversed if $\theta \in [\pi,2\pi]$.

Having become familiar with the unperturbed shape of the isovelocity curves, 
we are now ready to add planets.
To do so, in the following section we apply the planet velocity perturbations to the unperturbed Keplerian velocity field.

\subsection{Analytic kinks}

We use the linear and nonlinear velocity perturbations, computed respectively in \cref{linear} and \cref{nlvelocity}, in order to predict the shape of the kinks arising from the planetary wake. In the presence of a planet, the velocity field at each point of the disc is given by the unperturbed Keplerian velocity (\ref{velocityd}) plus the radial and azimuthal velocity perturbations $u$ and $v$:
\begin{equation}
    \begin{aligned}
    \mathbf{v}(r,\varphi) & = \bigl(v_\textrm{K}(r) + v(r,\varphi)\bigr)\mathbf{e}_\varphi + u(r,\varphi) \mathbf{e}_r \\
    & = \bigl(u\cos \varphi - (v+v_\textrm{K})\sin\varphi \bigr)\mathbf{e}_x + \bigl(u\sin \varphi + (v + v_\textrm{K})\cos \varphi \bigr)\mathbf{e}_y.
    \end{aligned}
    \label{withplanet}
\end{equation}
The application of these perturbations is split into linear and nonlinear regimes. In the linear regime for $u$ and $v$ we used the dimensionless velocity perturbations computed in \cref{linear}, rescaled by a factor $c_\textrm{p}(M_\textrm{p}/m_\textrm{th})$. These perturbations were applied to the disc grid points inside the linear box domain around the planet.
In the nonlinear regime $u$ and $v$ were computed using equations (\ref{unl}) and (\ref{vnl}), where the terms $\Lambda _\textrm{p}f_u$ and $\Lambda _\textrm{p}f_v$ are given by (\ref{pwlu}) and (\ref{pwlv}), choosing $\delta=1$ and $q = 1/4$. For the term $\chi$ we used the numerical solution of Eq. (\ref{burgers}) if $t<t_\infty$ and the analytical N-wave solution (\ref{Nwave}) for larger $t$.

\begin{figure}
    \centering
    \includegraphics[scale=0.3]{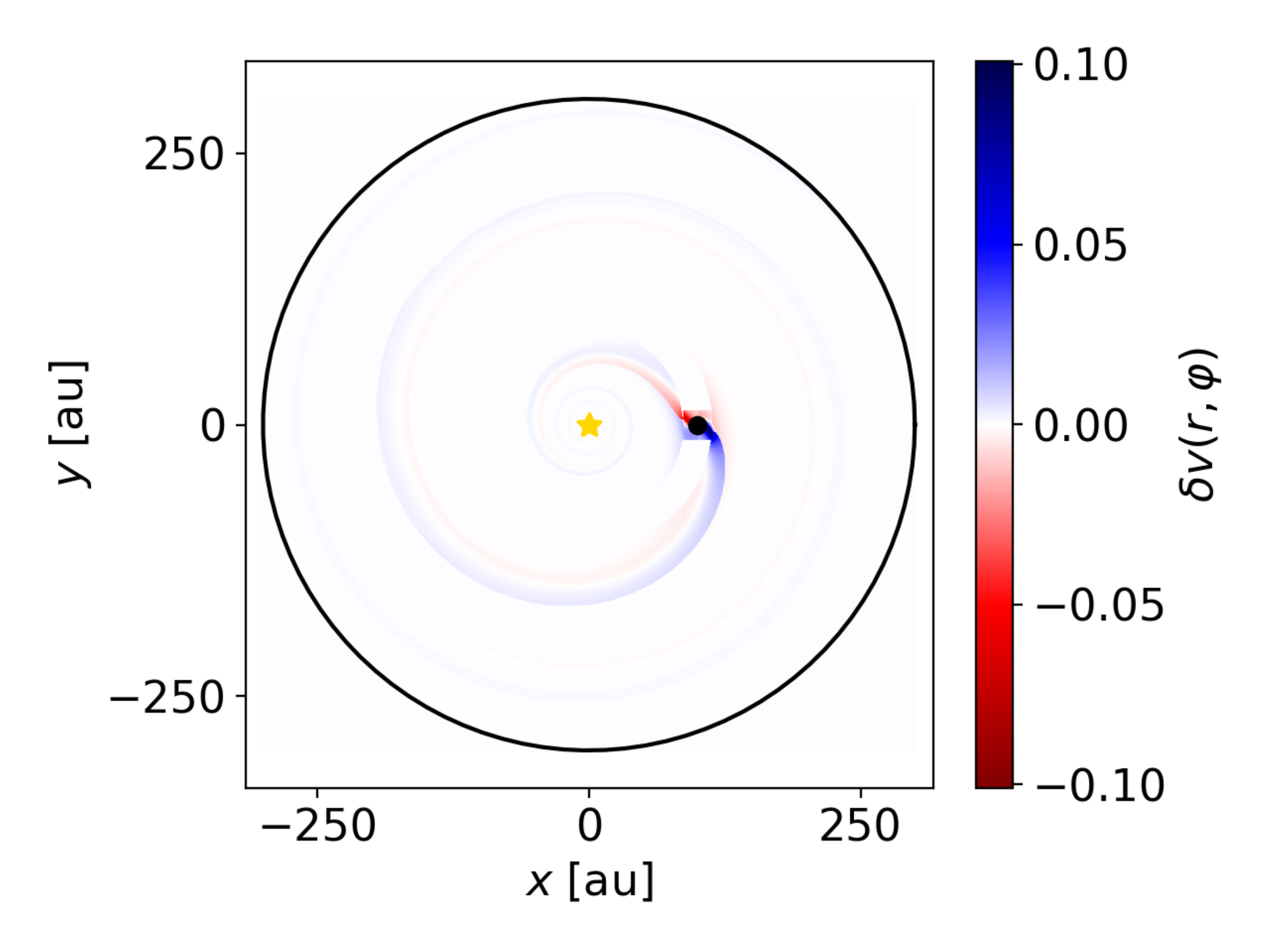}
    \caption{The velocity perturbation in the whole disc is non-vanishing along the spiral density wake. The black circle denotes the planet location.}
    \label{ideltav}
\end{figure}

In order to visualize how the disc velocity field is altered by the presence of the planet, we considered the quantity $\delta v \equiv (|\mathbf{v}|-|\mathbf{v}_0|)/|\mathbf{v}_0|$ where $\mathbf{v}$ and $\mathbf{v}_0$ are the disc velocity in the presence and in the absence of the planet (Eqs. \ref{withplanet} and \ref{velocityd} respectively).
Figure \ref{ideltav} shows the contour plot of $\delta v$ for a system with $M_\star = M_\odot$, $R_\textrm{disc}=300$ au, $\gamma = 5/3$ and $h_\textrm{p}/r_\textrm{p}=c_\textrm{p}/v_\textrm{K}(r_\textrm{p}) = 0.1$, which implies a thermal mass (\ref{unitm}) of $m_\textrm{th}=0.7 M_\textrm{Jupiter}$.
The planet has a mass
$M_\textrm{p}=M_\textrm{Jupiter}$ and is located at $(r_\textrm{p}=100\,\textrm{au}, \varphi _\textrm{p}=0)$.
From this Figure we observe that
the velocity perturbation along the planetary wake undergoes an abrupt sign reversal across the planet. This feature of the wake was suggested by
\cite{Casassus19} to infer the presence and location of a planet in the system HD 100546 through the corresponding ``Doppler flip'' in the first moment map.

The kinks in the channel maps arise due to the variation of the component along $\hat{n}$ of the disc velocity that occurs in presence of a planet. For an unperturbed disc, the gas velocity projected on $\hat{n}$ is given by (\ref{vnk}). Instead, if the disc hosts a planet, the dot product of Equation~(\ref{withplanet}) and $\hat{n}$ yields
\begin{equation}
    v_n(r,\varphi)= \sin \theta \bigl[\bigl(v_\textrm{K}(r)+v(r,\varphi)\bigr)\sin \varphi - u(r,\varphi)\cos \varphi \bigr].
    \label{vn}
\end{equation}
As in the unperturbed case (see \cref{channels}), the analytical channel map with channel velocity $v_\textrm{ch}$ is produced by ``painting'' the grid points such that $v_n(r,\varphi)\in [v_\textrm{ch}-\Delta v,v_\textrm{ch}+\Delta v]$.

\begin{figure}
    \centering
    \includegraphics[scale=0.3]{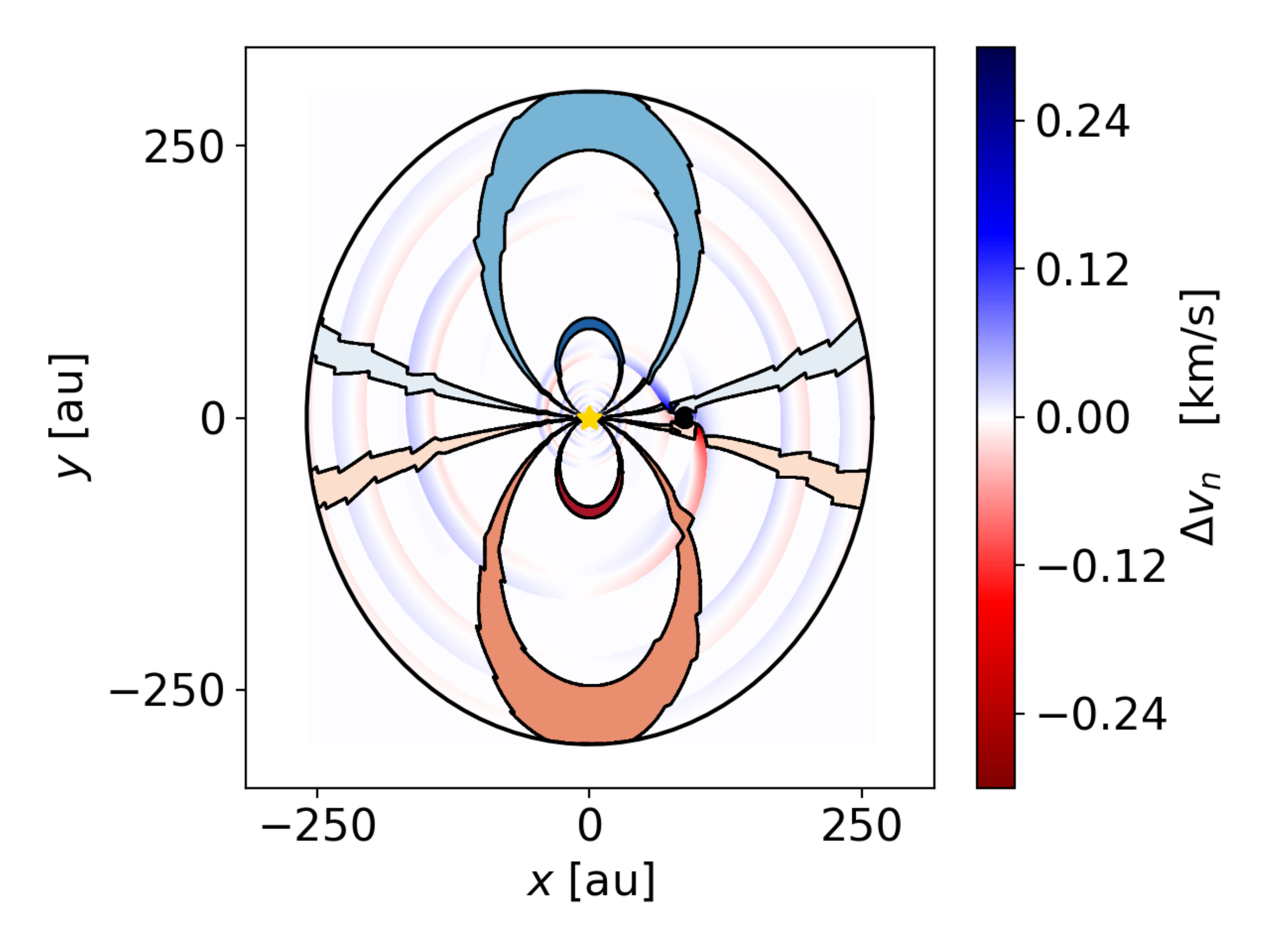}
    \caption{Analytical channel maps of Fig.\ref{iCM1} for a disc hosting a planet, with $\theta = \pi/6$ and channel velocity resolution $\Delta v=0.05$ km/s. In the figure is also shown the countour of $\Delta v_n$ (Eq. \ref{Dvn}). The kinks appear in the whole disc almost every time a channel map crosses the spiral planetary wake.}
    \label{iCM2}
\end{figure}

Figure~\ref{iCM2} shows the resulting channel maps.
In the background of this figure we also show the quantity
\begin{equation}
 \Delta v_n \equiv v_n - v_{n,0} = \sin \theta \bigl[v(r,\varphi)\sin \varphi - u(r,\varphi)\cos \varphi \bigr],
 \label{Dvn}
\end{equation}
We observe that the velocity perturbations, once added to a Keplerian velocity field, give rise to the observed kinks. These are created when the planet wake crosses a velocity channel.

\subsection{Velocity damping} \label{SecVisc}
Figure \ref{iCM2} shows that the theoretical kinks are present even in those channel maps that cross the wake in regions of the disc far away from the planet. This is in conflict with the observational and numerical evidence showing that kinks are localized near the planet \citep{Pinte18,Pinte19}. While this may just be due to the finite beam resolution that would smear out the smaller kinks far from the planet, we also discuss here the hypothesis that this excess of kinks is due to a lack of damping of the analytical velocity perturbations as the wake travels away from the planet. Viscosity is known to damp density and velocity perturbations as they propagate across the disc (\citealt{S84}) and we thus apply an exponential damping to the velocity perturbations in order to recover the local behavior in the theoretical kinks.

 \cite{S84} showed that a linear density wave excited at the $m$-th Lindblad resonance $r_\textrm{r}$ undergoes viscous damping by a factor
\begin{equation}
    \exp \Bigl(-\int_{r_\textrm{r}}^r k_\textrm{I} \textrm{d}r' \Bigr),
    \label{exp}
\end{equation}
where
\begin{equation}
    k_\textrm{I} = -\epsilon\frac{(7\nu_\textrm{s}/3)\kappa k_\textrm{R}^2}{2\pi G\Sigma _0 - 2 c_0^2|k_\textrm{R}|},
    \label{ki}
\end{equation}
where $\nu _\textrm{s}$ is the shear viscosity, $\epsilon = \pm 1$ at inner/outer Lindblad resonances and $\kappa$ is the epicyclic frequency. The real part of the wave number $k_\textrm{R}$ satisfies the \citet{LS64} dispersion relation for tightly wound density waves in a two-dimensional gaseous disc, namely
\begin{equation}
    m^2\bigl(\Omega-\Omega_\textrm{p} \bigr)^2 = \kappa ^2 - 2\pi G\Sigma _0|k_\textrm{R}| + c_0^2k_\textrm{R}^2.
    \label{linshu}
\end{equation}
Equation (\ref{ki}) can be simplified with the following approximations. In the non-self gravitating limit we can neglect the term $2\pi G\Sigma _0$ in both equations (\ref{ki}) and (\ref{linshu}). In addition, if we take the Lindblad resonance index $m\gg 1$ and we use $\kappa = \Omega$ (as in Keplerian rotation), Eq. (\ref{linshu}) yields
\begin{equation}
    |k_\textrm{R}| = m\frac{|\Delta \Omega|}{c}.
    \label{kR}
\end{equation}
By replacing (\ref{kR}) in (\ref{ki}), assuming $\nu_\textrm{s} = \alpha h c$ \citep{SS73} and using (\ref{H}) , we obtain
\begin{equation}
    k_\textrm{I} \simeq -\epsilon \alpha m \frac{7|\Delta \Omega|}{6c_0}.
    \label{kim}
\end{equation}
Under these approximations the damping factor (\ref{exp}) for the velocity perturbations launched at the $m$-th Lindblad resonance can be written as
\begin{equation}
    \exp \Bigl(\epsilon m \alpha \frac{7}{6}\int _{r_\textrm{r}}^r \frac{|\Delta \Omega|}{c_0}\textrm{d}r'    \Bigr).
    \label{dampfac}
\end{equation}

The radial and azimuthal velocity perturbations $u$ and $v$
(Eqs. \ref{unl} and \ref{vnl} in the nonlinear regime and those computed in \cref{linear} for the linear regime) are the velocity perturbations associated with the planetary wake, which is the result of the constructive interference of the waves launched at several Lindblad resonances (\cref{wakeshape}). Each of the waves excited at Lindblad resonances has a different viscous damping factor due to the dependence of (\ref{dampfac}) on the index $m$.
Despite the fact that the perturbations $u$ and $v$ come from the superposition of waves with different damping factors, we introduce the disc viscosity in the model developed so far by applying to $u$ and $v$ a single exponential factor
\begin{equation}
    \mathscrsfs{D} = \exp \Biggl(-\frac{7\alpha m}{6h_\textrm{p}}\bigg|\int _{r_\textrm{p}}^r \big|(r'/r_\textrm{p})^{-3/2}-1\big|(r'/r_\textrm{p})^q \textrm{d}r'\bigg|\Biggr),
    \label{damp}
\end{equation}
which is obtained from Eq. (\ref{dampfac}) using $c_0 = c_\textrm{p}(r/r_\textrm{p})^{-q}$, $\Omega = (GM_\star/r^{3})^{1/2}$, Eq. (\ref{H}) and $r_\textrm{r} \simeq r_\textrm{p}$ (being $m \gg 1$).
Then, the velocity perturbations in the presence of viscous damping, $u_\textrm{d}$ and $v_\textrm{d}$, read
\begin{equation}
    \begin{aligned}
    u_\textrm{d}(r,\varphi) &= \mathscrsfs{D}(r)u(r,\varphi), \\
    v_\textrm{d}(r,\varphi) &= \mathscrsfs{D}(r)v(r,\varphi).
    \end{aligned}
\end{equation}
In our treatment the coefficient (\ref{damp}) appears as a term that mimics the effect of viscosity, whose strength can be tuned by varying the parameter $\alpha m$.

\begin{figure}
    \centering
    \includegraphics[scale=0.3]{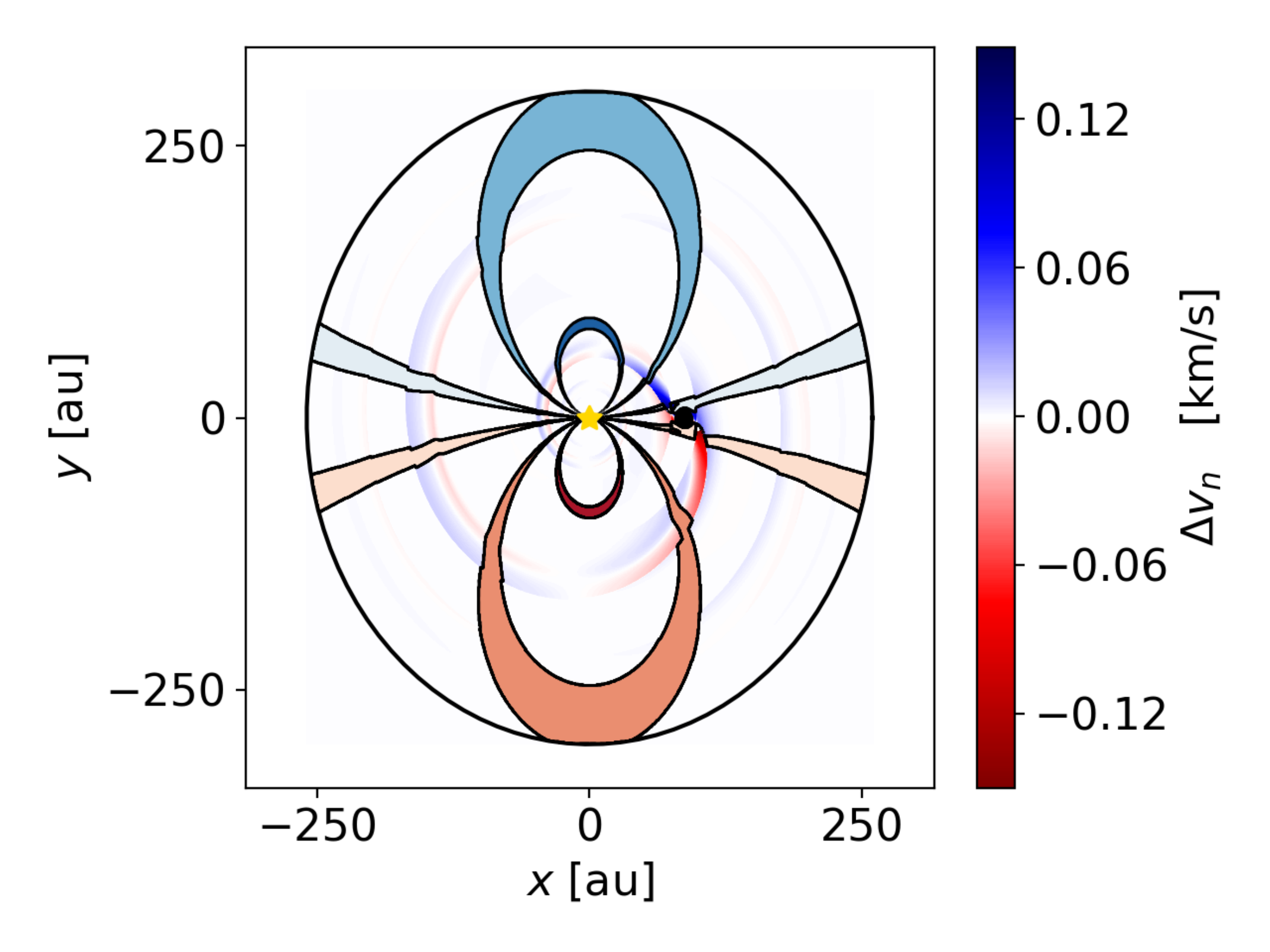}
    \caption{Analytical channel maps of Fig. \ref{iCM1} for a disc hosting a planet, with $\theta = \pi/6$ and channel velocity resolution $\Delta v=0.05$ km/s. In the figure is also shown the countour of $\Delta v_n$ (Eq. \ref{Dvn}). The kinks appear only in the vicinity of the planet due to a viscous damping with $\alpha m = 0.2$.}
    \label{iCM3}
\end{figure}

Figure \ref{iCM3} shows how the analytical kinks in the channel maps of Figure~\ref{iCM2} are modified in presence of viscous damping.
In order to evaluate (\ref{damp}) we assumed a sound speed profile index $q = 1/4$ and we tuned $\alpha m$ such that the kinks were sufficiently damped after about a half-winding of the wake, as it occurs in observations and numerical simulations. This led us to set $\alpha m = 0.5$.

\begin{figure*}
  \includegraphics[scale=0.345,trim=-2cm  -3.1cm 0 0]{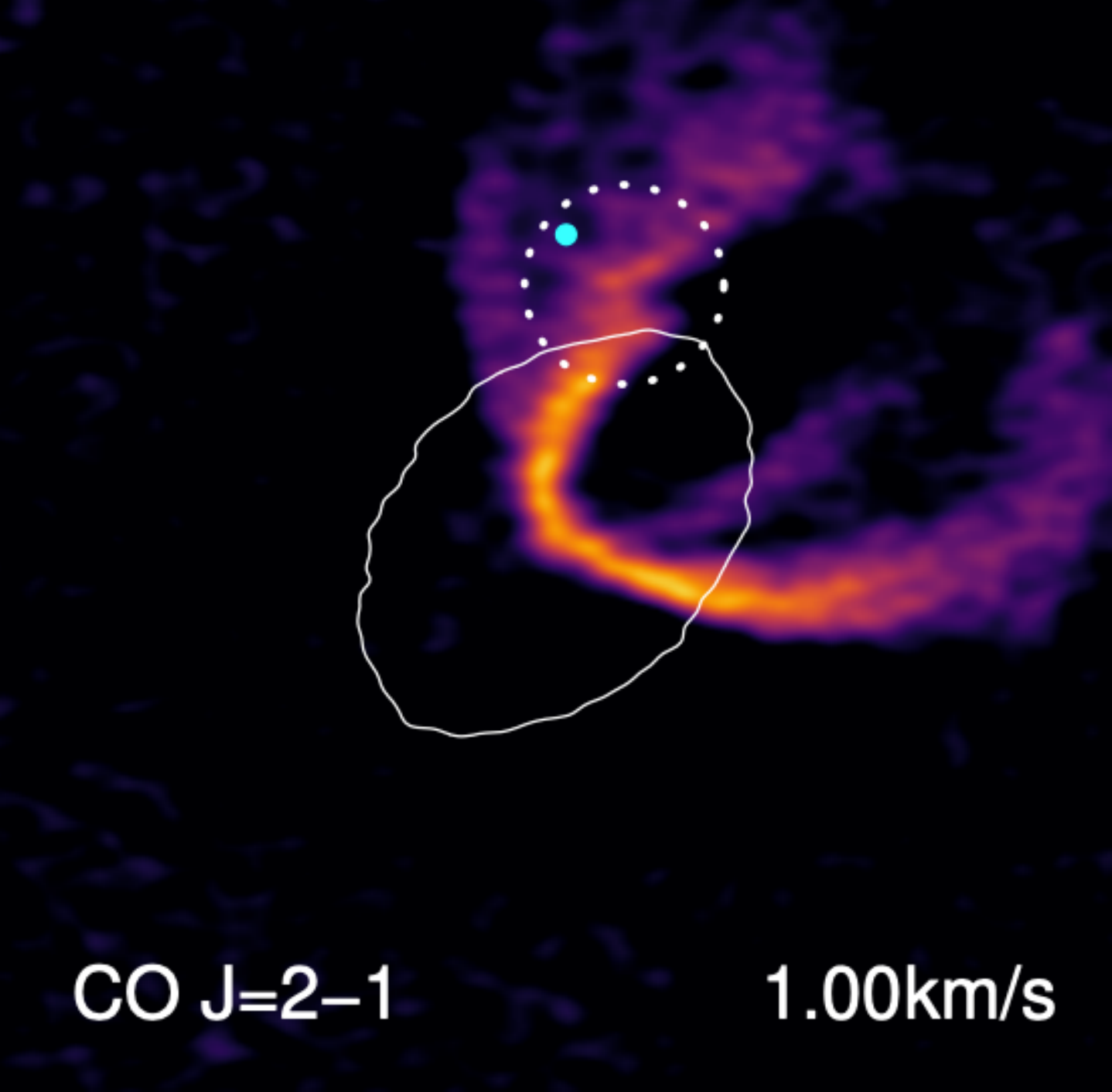}
  \includegraphics[scale=0.3, trim=-2cm 0 0 0]{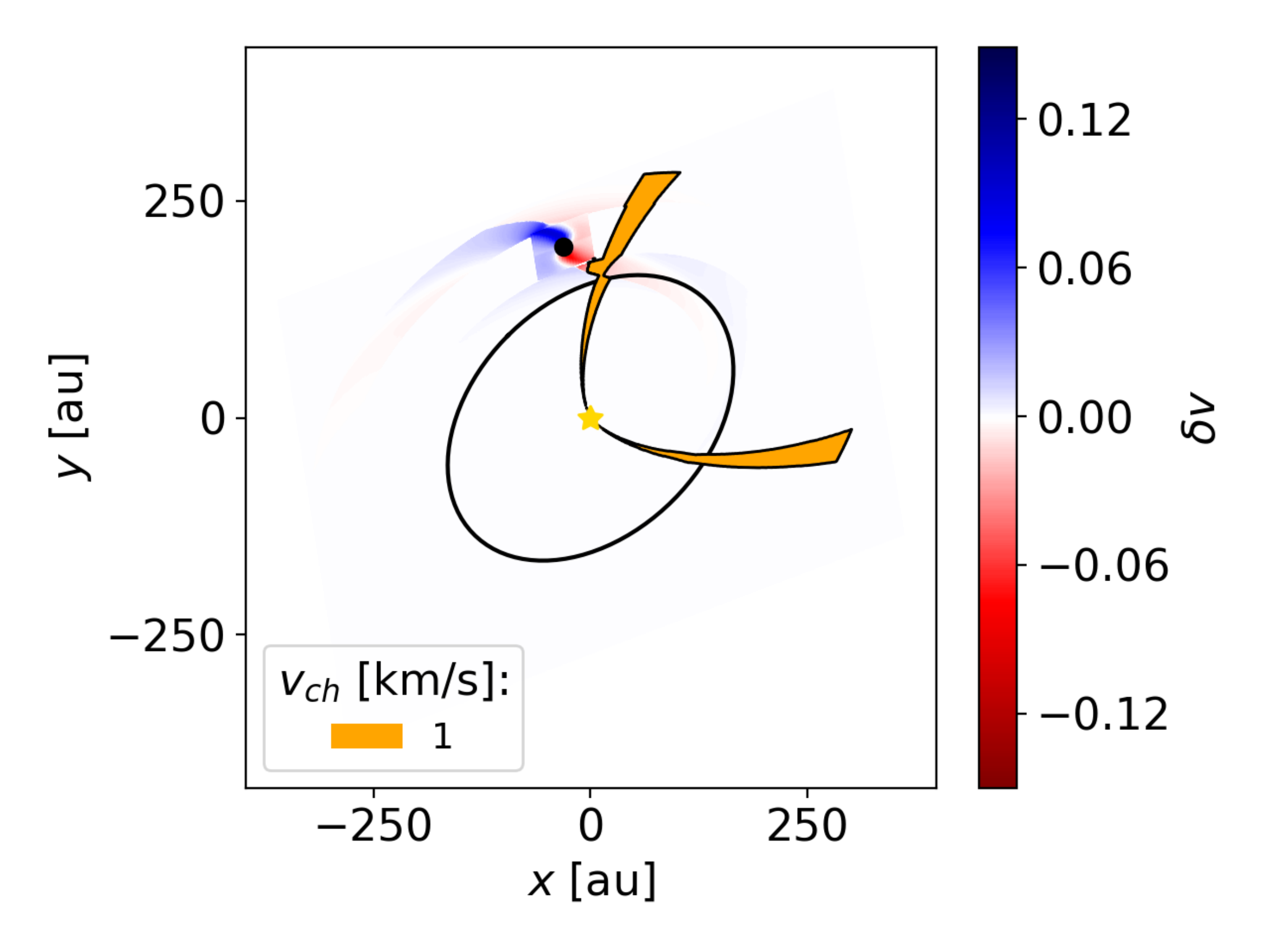}
  \caption{The left panel (Credit: \citealt{Pinte18}) shows the channel map of $^{12}$CO line emission at +1 km/s from the systemic velocity of the source HD 163296. The kink in the emission is highlighted by the dotted circle. The potential planet location is marked by a cyan dot.  In the right panel the corresponding theoretical channel map is shown. The contour of $\delta v$ appears in blue and red. In both images the ellipses represents the edge of the dust disc.}
  \label{iPinte18}
\end{figure*}

We now use our analytic model for the channel maps to make a comparison with the the channel map of the disc surrounding HD 163296
studied in \cite{Pinte18}.
Figure~\ref{iPinte18} shows that the basic morphology of the kink is the same as in the observation reported in \cite{Pinte18} when compared to the corresponding theoretical channel map created using the system parameters as derived by \cite{Pinte18}, based on a comparison with their hydrodynamical simulations: a planet mass $M_\textrm{p} = 2 M_\textrm{Jupiter}$ in orbit at $r_\textrm{p} = 270$ AU around a $1.9 M_\odot$ star in a disc with inclination $i = 45^{\circ}$, PA $= 33.75 ^{\circ}$, aspect ratio $h_\textrm{p}/r_\textrm{p} = 0.1$, power-law profiles $\Sigma _0= \Sigma _\textrm{p}(r/r_\textrm{p})^{-1}$, $c_0 = c _\textrm{p}(r/r_\textrm{p})^{-1/4}$ and adiabatic index $ \gamma = 5/3$.

Here we provided a model for the wave damping due to viscosity, but other mechanisms, whose discussion is beyond the aim of this paper, can play a significant role in mitigating the planet induced perturbations, such as the presence of a finite cooling timescale \citep{Miranda2020}.

\subsection{Dependence of the kink amplitude on the system parameters}

Ultimately, we wish to understand how the kink amplitude can be used to measure the planet mass. For this we need to measure and interpret the ``distance'' of an isovelocity curve displaying a kink from its unperturbed shape.

\begin{figure}
    \centering
    \includegraphics[scale=0.6]{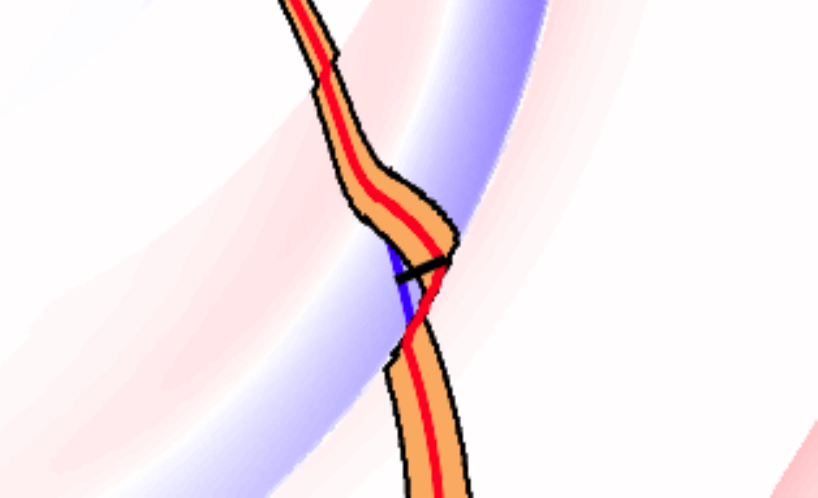}
    \caption{The channel map in orange shows a kink as it crosses the wake. The center of the channel map (red line) is displaced from its unperturbed position (blue line). The length of the black segment linking the two lines is the measure (\ref{kinkamplitude}) of the kink amplitude.}
    \label{iA}
\end{figure}

First, we introduce a quantitative measure for the kink amplitude. Let $I_1$ be the set of points in an isovelocity curve of a disc hosting a planet and $I_0$ the corresponding isovelocity curve for the unperturbed disc. If we consider a point $\mathbf{P}_1 \in I_1$, its distance from the curve $I_0$ is given by
\begin{equation}
   \min _{\mathbf{P}_0 \in I_0}\left[d_\textrm{E}(\mathbf{P}_0,\mathbf{P}_1) \right],
   \label{dist}
\end{equation}
where $d_\textrm{E}(\mathbf{P}_0,\mathbf{P}_1) = [(x_{\mathbf{P}_1}-x_{\mathbf{P}_0})^2+(y_{\mathbf{P}_1}-y_{\mathbf{P}_0})^2]^{1/2}$.
Then, we define the kink amplitude $\mathscrsfs{A}$ as the maximum of the quantity (\ref{dist}) over the points $\mathbf{P}_1 \in I_1$:
\begin{equation}
    \mathscrsfs{A} = \max _{\mathbf{P}_1 \in I_1}\Bigl[  \min _{\mathbf{P}_0 \in I_0}\Bigl( d_\textrm{E}(\mathbf{P}_0,\mathbf{P}_1) \Bigr) \Bigr].
    \label{kinkamplitude}
\end{equation}
This quantity measures the length of a segment that connects $I_1$ and $I_0$, hence $\mathscrsfs{A}$ has the dimension of a length (Figure~\ref{iA} shows the geometrical interpretation of (\ref{kinkamplitude})).
With this tool we studied the dependence of the kink amplitude on the planet mass and on the channel velocity of the isovelocity curve.

\begin{figure}
    \centering
    \includegraphics[scale=0.3]{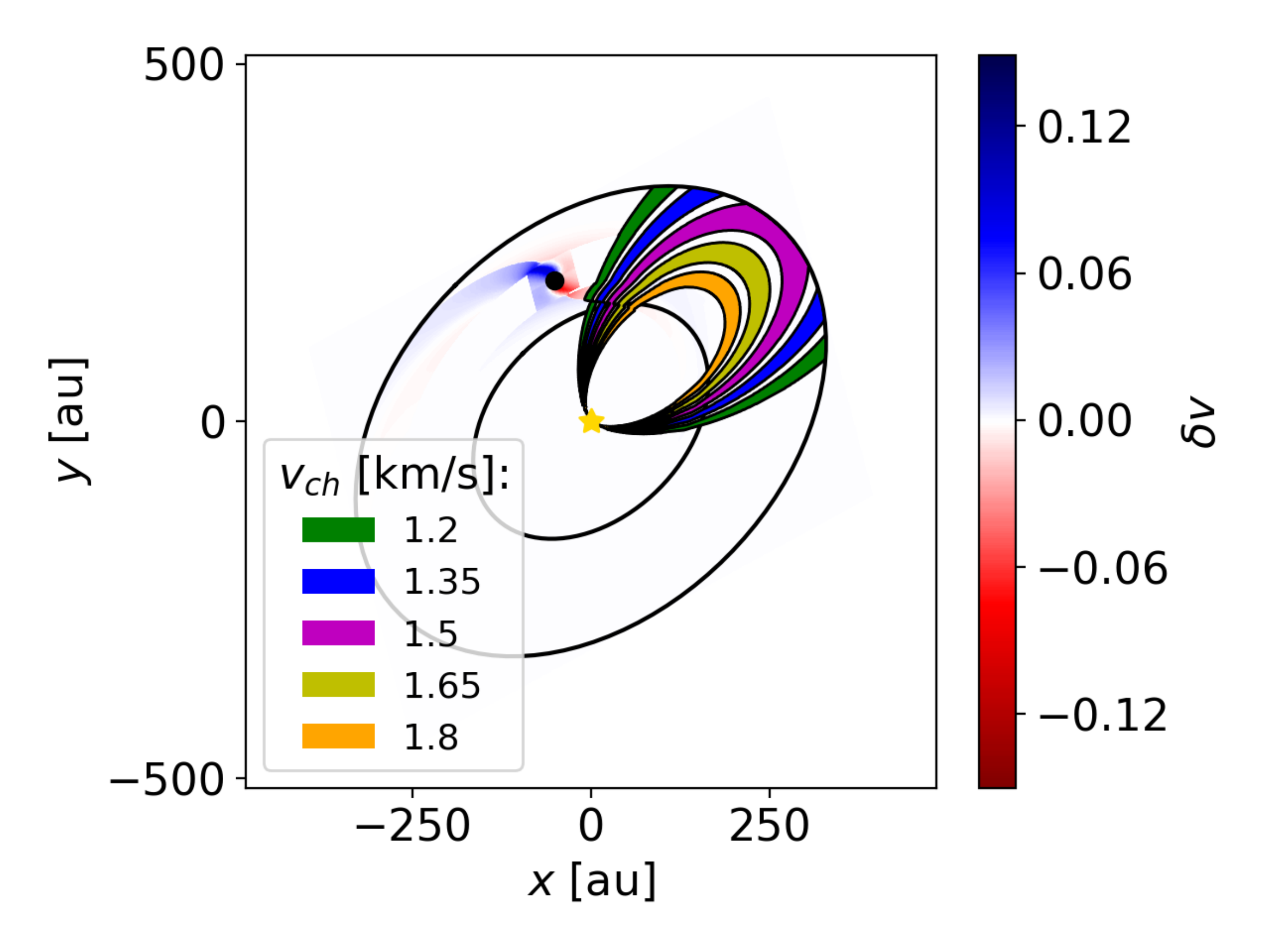}
    \caption{The channel maps of the system HD 163296 considered to study the kink amplitude dependence on the planet mass and channel velocity. The inner ellipses represents the dust disc edge whereas the outer ellipses marks the edge of the gaseous disc.}
    \label{ichannels}
\end{figure}

We adopted the system parameters of the source HD 163296 studied in \citet{Pinte18}.  We considered the channel velocities $v_\textrm{ch}/(\textrm{km/s})\in \mathcal{V}=\{ 1.2, 1.35, 1.5, 1.65, 1.8 \}$, chosen such that the corresponding channel maps, shown in Fig. \ref{ichannels}, cross the inner wake (the branch inside the planet orbital radius).
For each channel velocity in $\mathcal{V}$ we computed the kink amplitude $\mathscrsfs{A}$ of the kink nearer to the planet, for different planet masses $M_\textrm{p}/M_\textrm{Jupiter} \in \{0.5, 1, 1.5, 2, 2.5, 3, 3.5, 4\}$.

\begin{figure}
    \centering
    \includegraphics[scale=0.5]{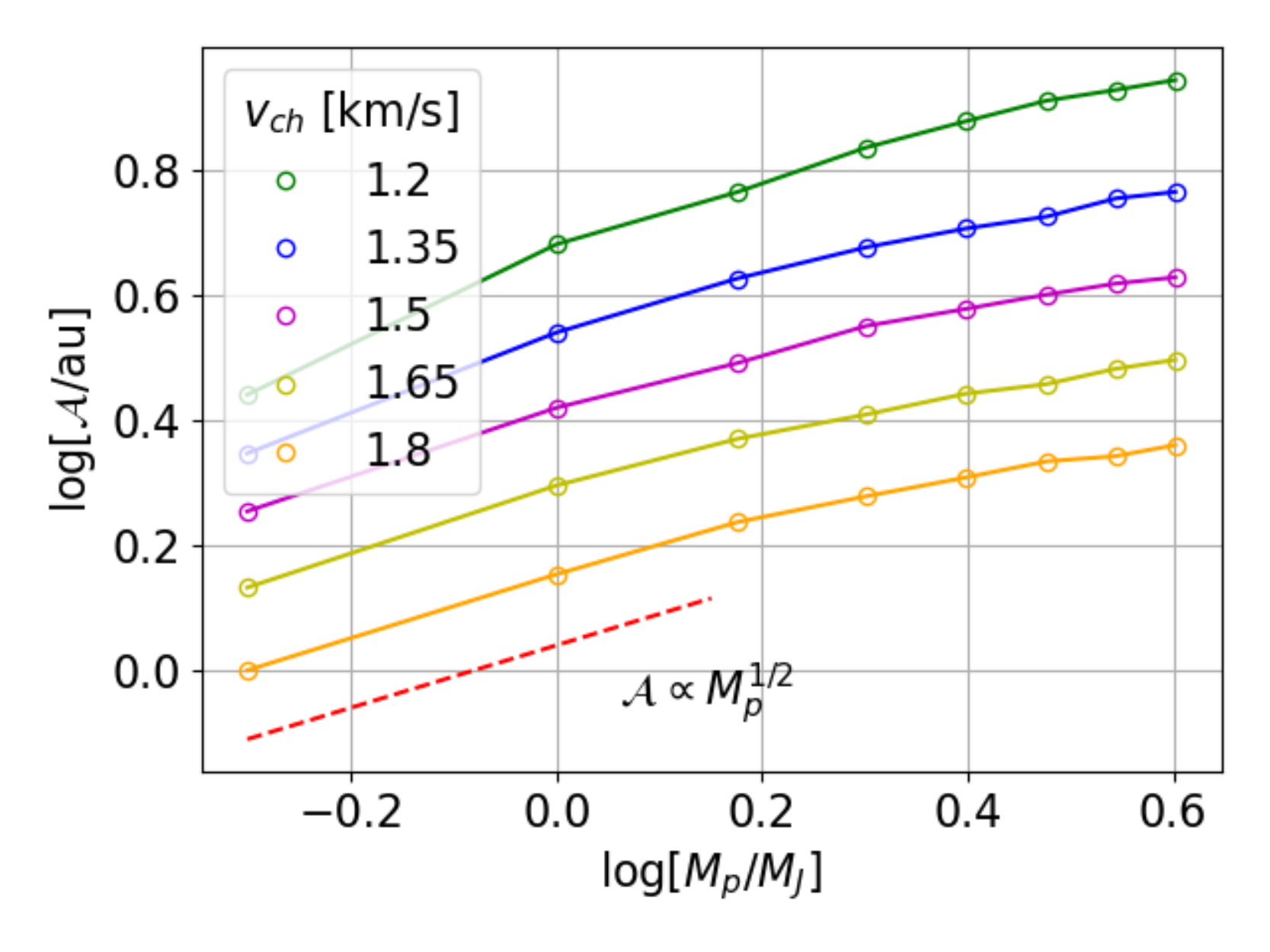}
    \caption{Kink amplitude dependence on the planet mass for the channel maps of Fig. \ref{ichannels}.}
    \label{ikinkvsmp}
\end{figure}

Figure~\ref{ikinkvsmp} shows the results. We observe that for a fixed planet mass, the kink amplitude grows as channels move closer to the planet (see Figure~\ref{ichannels}). This is expected since near the planet the interaction with the disc is stronger and then the kinks arising from the resulting velocity perturbations are likely to be more prominent.
In other words, the kink amplitude becomes larger as $v_\textrm{ch} \to v_\textrm{ch,p}$, where $v_\textrm{ch,p}=v_\textrm{K}(r_\textrm{p})\sin \theta \sin \varphi _\textrm{p}$ is the channel velocity corresponding to the line of sight projection of the planet Keplerian motion. For the system parameters shown in Figs. \ref{ichannels} and \ref{ikinkvsmp}, $v_\textrm{ch,p} \simeq 0.68$ km/s.

Moreover, for a fixed channel velocity, the curves in Figure~\ref{ikinkvsmp} show that the kink amplitude increases with  planet mass.
Indeed, a more massive planet generates a stronger velocity perturbation and hence a greater $\Delta v_n$ (Eq. \ref{Dvn}) which implies a larger deviation of the isovelocity curve from its unperturbed shape. In addition, the proportionality between $\mathscrsfs{A}$ and $M_\textrm{p}$ weakens as we move away from the planet (curves from top to bottom), approaching a $\mathscrsfs{A}\propto M_\textrm{p}^{1/2}$ dependence (at least for the lowest considered masses). This behavior is in agreement with the velocity perturbations dependence on the planet mass,
which is linear near the planet, becoming less steep in the nonlinear regime where the height of the N-wave lobes asymptotically approach a $\propto M_\textrm{p}^{1/2}$ scaling.
We found similar results
for the dependence of $\mathscrsfs{A}$ on $v_\textrm{ch}$ and $M_\textrm{p}$
 when considering channel maps crossing the outer wake.


Finally, there is another geometrical factor that affects the kink amplitude, related to the position of the kink with respect to the  major-axis of the disc.
Namely, the kink amplitude depends on the quantity $\Delta v_n$. Let us consider the simple case where the disc major axis is the $y$-axis. From Eq. (\ref{Dvn}) we see that when the planet lies on the $x$-axis ($\varphi _\textrm{p}=0$ or $\varphi _\textrm{p}=\pi$) the amplitude of a kink in its neighbourhood will be related to $\Delta v_n \simeq \mp u\sin \theta$. On the other hand, when the planet is on the $y$-axis ($\varphi _\textrm{p} = \pm \pi/2$) we have $\Delta v_n \simeq \pm  v\sin \theta$.
In addition, from Equations~(\ref{unl}) and (\ref{vnl}) it follows that the azimuthal velocity perturbation $v$ is smaller than $u$ by a factor $h_\textrm{p}/r_\textrm{p}$. Therefore $\Delta v_n$ will be larger when the planet is situated near the  minor axis, as can be seen in
Figure~\ref{iCM2}, where $\Delta v_n$ appears to be smaller near the $y$-axis.
As a consequence, also the kink amplitude will be larger near the minor-axis.
Confirmation of this feature can be found in \cite{Perez2018}, where hydrodynamic simulations of protoplanetary discs revealed that the ``wiggles'' in the first moment maps caused by the planet-induced velocity perturbations were more pronounced when the planet was located near the disc minor-axis.

\section{Discussion}\label{discussion}
In this paper we have developed a semi-analytic theory for the `kink' perturbation induced in channel maps of line emission by a planet embedded in a circumstellar disc. The basic procedure for predicting the velocity perturbation caused by the planet is as follows:
\begin{enumerate}
    \item Solve for the linear structure of the wake in the vicinity of the planet in the shearing-sheet approximation using a Fast Fourier Transform, similar to the procedure in \citetalias{Goodman01}
    \item Use this as the boundary condition for computing the non-linear structure of the wake by solving the corresponding Burger's equation (\ref{burgers}) numerically, as in \citetalias{Rafikov02}.
    \item Relate the radial and azumuthal velocity perturbation amplitudes to the density perturbation via (\ref{unl}) and (\ref{vnl}), multiplying these by an optional damping term (\ref{dampfac}).
    \item Add the perturbations to the unperturbed Keplerian velocity profile (\ref{withplanet}) and project at the desired position angle and inclination to find the line of sight velocity component (\ref{vn}).
    \item Compute synthetic channel maps, comparing the amplitude of the kink defined via (\ref{kinkamplitude}) to that measured from the observations.
\end{enumerate}

There are three main limitations to our analytic model. First, the theory of \citetalias{Goodman01} and \citetalias{Rafikov02} is only technically valid for non-gap-opening planets with mass below the thermal mass (\ref{unitm}), which is $\approx 0.7$ M$_\textrm{Jup}$ for a disc with an aspect ratio $h_\textrm{p}/r_\textrm{p} = 0.1$ around a solar mass star. This may be problematic given inferred planet masses of 2--3 $M_\textrm{Jup}$ for the kinks observed by \citet{Pinte18,Pinte19}. Planet masses above the thermal mass cause a deviation from \citetalias{Goodman01} and \citetalias{Rafikov02} theory because the linear and non-linear evolution of the waves can no longer be separated, as the waves launch already in a non-linear fashion. Furthermore, planets above the thermal mass should open a gap, which changes the assumed background surface density. Mitigating this is that the kink is due to the non-linear part of the wake located outside the planet gap, where the azimuthally-averaged disc surface density is relatively undisturbed. Indeed the comparison with the observations in Figure~\ref{iPinte18} suggests that these effects are not too severe. However, a more detailed comparison of our analytic predictions with 3D hydrodynamical simulations would be useful to quantify this, similar to the comparison with \citetalias{Rafikov02} already performed in \citet{Zhu15}. A possible improvement would be to use a prescribed gap profile for the background surface density \citep[e.g.][]{Crida06}, but this would still leave the problem of overlapping linear/non-linear regimes.

Second, the analytic theory has only been developed in two dimensions, meaning that it assumes no vertical motion. While the vertical motions induced by the planet are smaller than the radial and azimuthal velocity perturbations \citep[see e.g. supplementary material in][]{Pinte19}, they are not negligible. We have also assumed a thin disc with a sound speed that is prescribed as a function only of radius, which excludes the possibility of vertical temperature stratification --- causing refraction of the waves producing the planet wake --- and buoyancy, which can induce additional vertical motion \citep{Bae21}. Also, the CO emitting layer that is commonly used to detect planetary kinks is located at $\approx 3-4$ pressure scale-heights from the midplane, thus inducing projection effects that are not taken into account in 2D, as mentioned in \cref{channels}.

Third, the analytic theory also assumes a single embedded planet on a circular orbit in the plane. Hence the kink signature may be different for a planet on an inclined or eccentric orbit. This limits the application to a subclass of observed discs. For example, there are at least two planets imaged in the PDS 70 disc \citep{Keppler19,Haffert19} which models suggest are on eccentric orbits \citep{Bae19,Muley19,Toci20}.

Finally, we have also not considered the effects of noise, radiative transfer, chemistry or other real-world complications in producing our synthetic line emission maps.

In step iv) of the above we have assumed that the kink seen in the channel maps is entirely caused by the planet wake. That this succeeds in reproducing the observed kink morphology suggests that this hypothesis is indeed correct and that the kink is caused by the planet wake and not by a circumplanetary disc \citep{Perez15} or radial inflow into a gap \citep{Teague18}. The caveat is that --- in contrast with observations and the results of 3D hydrodynamical simulations \citep{Pinte18,Pinte19} --- we found that additional, albeit weaker, kinks should be visible further away from the planet, each time the channel maps cross the wake (Figure~\ref{iCM2}). One possible reason for this in the analytic theory is the absence of additional damping caused by viscosity or other small scale motion. We hence extended this theory to include the effect of additional damping as would be caused by a Shakura-Sunyaev $\alpha$-viscosity. Adding such damping produced results more in agreement with previous findings, where we require $\alpha m \approx 0.5$, where $m$ is the order of the Lindblad resonance dominating the wake. The strongest angular momentum exchange between the planet and disc occurs at the Lindblad resonances with $m\sim r_\textrm{p}/h_\textrm{p} = 10$ \citep{Rafikov02}, suggesting $m\sim 10$ but this would imply excess viscosity compared to the 3D simulations and compared to observational constraints on small scale motions in protostellar discs by \citet{Flaherty18,Flaherty20}.
Note that \citet{Dong2011} showed numerically that viscous damping of nonlinear waves excited by sub-thermal planet masses can be significant even at low viscosity $\alpha \sim 10^{-4}$, which may indicate that the damping prescription adopted by us (Eq. \ref{damp}) underestimates the actual non-linear damping, reducing the need for high viscosity.
Another possibility for the lack of the additional kinks further away from the planet
is that wake damping is caused by thermodynamic effects, as discussed by \citet{Miranda2020}. A final possibility is that the secondary kinks might simply be washed out by observational effects, including a finite beam size and spectral resolution, and that CO emission is not produced in the midplane. We intend to explore further all the issues above in subsequent papers in this series.

While we have demonstrated that the analytic theory for planet-disc interaction indeed produces kinks with roughly the correct morphology compared to the observations, our aim in subsequent papers in this series will be to develop this into a parameterised and automated fitting technique for measuring the planet mass with error bars. Further worthwhile effort could be spent in extending the theory to three dimensions, to multiple planets on eccentric or inclined orbits, and perhaps also to considering not only the gas but also the dust component \citep[e.g.][]{Dipierro18}.


\section{Conclusions}
\label{conclusions}
Disc kinematics offers a new, potentially very powerful technique to detect newly born planets that are still embedded in their natal disc. Such perturbations result in peculiar ``kinks'' in the channel maps of different gas species, such as CO, when observed at high spatial and spectral resolution with the Atacama Large Millimetre Array (ALMA). A limitation of this method has been the absence of a general theory relating the observed perturbations to the planet properties, meaning that the observed channel maps are usually compared to computationally expensive hydrodynamical simulations, preventing a proper fitting for the relevant system parameters.
Here, we have provided such an analytic theory, by extending previous analyses of the disc response to the presence of a massive planet both in the linear and in the non-linear regime. We:
\begin{enumerate}
\item confirm that the observed kinks are
consistent with the planet-induced wake;
\item provide quantitative relations between the kink amplitude and planet properties;
\item showed how to extend the theory to include the effects of damping, such as that caused by disc viscosity, which may be needed in order to have localized kinks that appear only in a limited range of velocity channels.
\end{enumerate}
Our semi-analytic model is intended to form the basis of a more quantitative analysis of the gas kinematics in planet-hosting discs.



\section*{Acknowledgements}
We thank the referee, Ruobing Dong, for useful comments that helped us to improve the presentation.
DJP and CP acknowledge funding from the Australian Research Council via FT130100034, DP180104235 and FT170100040. We received funding from the European Union’s Horizon 2020 research and innovation program under the Marie Sk\l{}odowska-Curie grant agreement NO 823823 (RISE DUSTBUSTERS project). We thank Roman Rafikov, Cathie Clarke, Giovanni Rosotti and Valentin Christiaens for interesting discussions.

\section*{DATA AVAILABILITY STATEMENT}
The code used to compute velocity perturbations and  produce figures is publicly available at \url{https://github.com/fbollati/Analytical\_Kinks}.




\bibliographystyle{mnras}
\bibliography{kinks} 







\bsp	
\label{lastpage}
\end{document}